\documentclass[conference,10pt]{IEEEtran}
\IEEEoverridecommandlockouts

\usepackage{cite}
\usepackage{amsmath,amssymb,amsfonts}
\usepackage{algorithm}
\usepackage{algcompatible}
\algblockdefx{FORP}{ENDFORP}[1]%
  {\textbf{for}#1 \textbf{do in parallel}}%
  {\textbf{end for}}
\usepackage{graphicx}
\usepackage{textcomp}
\usepackage{xcolor}
\usepackage{tikz}
\usepackage{pgfplots}
\usepgfplotslibrary{fillbetween}
\usepackage[binary-units=true]{siunitx}

\begin{document}

\title{A Fully-Integrated 5mW, 0.8Gbps Energy-Efficient Chip-to-Chip Data Link for Ultra-Low-Power IoT End-Nodes in 65-nm CMOS}

\author{
Hayate~Okuhara,
Ahmed~Elnaqib,
Martino Dazzi,
Pierpaolo Palestri,
Simone~Benatti,
Luca~Benini,
Davide~Rossi
%\thanks{Manuscript received xx xx, 2020; revised xx xx, 2020}%
\thanks{This work was supported in part by the WiPLASH (Architecting More Than Moore – Wireless Plasticity for Heterogeneous Massive Computer Architectures) project founded from
the European Union’s Horizon 2020 research and innovation program under Grant Agreement No. 863337.}%
\thanks{© 2021 IEEE.  Personal use of this material is permitted.  Permission from IEEE must be obtained for all other uses, in any current or future media, including reprinting/republishing this material for advertising or promotional purposes, creating new collective works, for resale or redistribution to servers or lists, or reuse of any copyrighted component of this work in other works.}%
\thanks{IEEE Transactions on Very Large Scale Integration (VLSI) Systems}
\thanks{DOI: 10.1109/TVLSI.2021.3108806}
}
\maketitle

\begin{abstract}
The increasing complexity of Internet-of-Things (IoT) applications and near-sensor processing algorithms is pushing the computational power of low-power, battery-operated end-node systems. This trend also reveals growing demands for high-speed and energy-efficient inter-chip communications to manage the increasing amount of data coming from off-chip sensors and memories. While traditional micro-controller interfaces such as SPIs cannot cope with tight energy and large bandwidth requirements, low-voltage swing transceivers can tackle this challenge thanks to their capability to achieve several Gbps of the communication speed at milliwatt power levels. However, recent research on high-speed serial links focused on high-performance systems, with a power consumption significantly larger than the one of low-power IoT end-nodes, or on stand-alone designs not integrated at a system level. This paper presents a low-swing transceiver for the energy-efficient and low power chip-to-chip communication fully integrated within an IoT end-node System-on-Chip, fabricated in CMOS 65nm technology. The transceiver can be easily controlled via a software interface; thus, we can consider realistic scenarios for the data communication, which cannot be assessed in stand-alone prototypes. Chip measurements show that the transceiver achieves 8.46x higher energy efficiency at 15.9x higher performance than a traditional microcontroller interface such as a single-SPI.

\end{abstract}

\section{Introduction}
\IEEEPARstart{P}{ushed} by the increasing complexity of near-sensor data analytics algorithms, the computational performance required by Internet of Things (IoT) end-nodes has increased dramatically in the last few years. Nowadays, near-sensor applications, such as Convolutional Neural Network (CNN) based image analysis and bio-metric processing, have to efficiently operate on large volumes of sensor data captured by microcontrollers, as well as application parameters such as weights of CNNs. To deal with this increasing complexity, state of the art SoCs (System on Chip) have already achieved performance in the order of several GOPS (Giga Operation per Seconds) within a power envelope in the order of a few mW, exploiting parallelism, Instruction Set Architecture (ISA) specialization, and domain-specific acceleration \cite{vega, mrwolf,samurAI}.

On the other hand, in modern embedded systems operating in the IoT domain, overcoming the limitations imposed by low chip-to-chip communication bandwidths and high energy consumption represent a major challenge. Conventional MCU peripherals, such as I2C and SPI, provide transfer data rates in the order of a few tens of Mbps, which are typically insufficient to satisfy the expected bandwidth and energy efficiency demand of the next-generation IoT applications. For example, according to the results reported in \cite{esweek2019}, the off-chip memory bandwidth required to perform MobileNetV2 inference \cite{MobileNet} at 10 FPS on an MCU is larger than 500 Mbps. Although there are some solutions which can reach this requirement (e.g., HyperBus or Octal SPI operating at fast frequencies) \cite{hyperRAM, apmemory,ssi}, their power consumption rapidly saturates the end-node power budgets.

Low voltage swing serial interfaces have the potential to solve the challenges mentioned above \cite{choi2015,choi2018,dazzi2018,Poulton2019,Pisati2020}. They utilize low-voltage swing signals at the physical layer, which enable data communications at a higher-rate and lower-power consumption than the existing digital interface. Indeed, pursuing the maximum bandwidth has been the key metric driving link design, and state-of-the-art links achieve tens or even hundreds of Gbps for high-performance applications \cite{Poulton2019,Pisati2020,Yousry2021}. In domains like the HPC, serial interfaces have already become an essential building block \cite{manticore}. Short-range serial links can also be used for chiplet-to-chiplet communication \cite{synopsys_ip}. Besides bandwidth improvements, extremely low energy-per-bit, below 1pJ/bit while maintaining a few Gbps of the rate has been achieved at a few Gpbs for low power applications \cite{choi2015,choi2018}. Hence, using energy efficient serial links in IoT edge devices is a promising direction  to mitigate I/O energy bottlenecks.

Although various research efforts have been reported in optimizing serial interfaces, their system-level integration, especially in edge-node microcontrollers, has not been extensively studied to the best of the authors' knowledge. Most of the designs are stand-alone and employ a verification module for laboratory experiments \cite{choi2018,Poulton2019,Pisati2020}.  In other words, these ignore the issues lying on system-level integration for real systems such as controller design to cooperate with an internal CPU, interface design to an SoC, link architecture optimization utilizing available resources in SoCs, and software interface. Unlike high-performance applications, these factors’ impacts cannot be ignored in IoT edge devices with limited resources.

From the observations above, the contributions of this paper are:

\begin{itemize}
\item We designed the first low-power serializer-deserializer link (SerDes) fully integrated into an open-source low-power microcontroller \cite{pulpissimo} fabricated in commercial 65-nm CMOS technology. Detailed architectural and micro-architectural information are provided. We implemented a communication protocol for a chip-to-chip data transfer with the SerDes, including software drivers. 
\item We evaluated the energy efficiency of the SerDes with real chip measurement results, deriving guidelines for its power management: we explored duty-cycled operation of the SerDes when full bandwidth is not required. 
\item We report on the trade-off between bandwidth and energy efficiency. The energy efficiency of the SerDes is finally compared with conventional digital peripherals widely adopted in microcontrollers such as SPI \cite{mrwolf,blackghost, intel2018}.
\end{itemize}

The SerDes achieves 8.46x higher energy efficiency at 793 Mbps compared to the case of a Single SPI operating at 50 Mbps. Moreover, even if we target a low bandwidth such as 10 Mbps with the SerDes, its efficiency is 8.58 higher than the SPI. Unlike other state-of-the-art links, these results are achieved with a complete interface integrated into a microcontroller.

The rest of this paper, based on preliminary circuit design exploration presented in \cite{Okuhara2020}, is organized as follows: In section \ref{sec:related_work}, we review the state-of-the-art of existing serial interfaces. Then the proposed system architecture is described in sections \ref{sec:overview} and \ref{sec:ser_des}. The inter-chip communication protocol is presented in section \ref{sec:chips_sync}. The chip implementation, and its evaluations are reported in sections \ref{sec:chip} and \ref{sec:measure}. Finally, we conclude this paper in section \ref{sec:conc}.

\section{Related work}\label{sec:related_work}

High-speed serial interfaces have been extensively studied over the last decades \cite{Yang1996,Dally1997,Lee2000,Kim2002,Bulzacchelli2006,Poulton2013,choi2015,choi2018,Poulton2019,Pisati2020,Yousry2021}.  Data rates reached 2.5 Gbps in the middle of the 1990s, as reported in \cite{Yang1996}. Nowadays, they exceed one hundred Gbps at a single lane, and tens of Tbps of aggregate bandwidth is achieved for many-lanes short-range interfaces \cite{Yousry2021}. To achieve such a high bandwidth, designers rely on advanced equalization techniques \cite{Dally1997,Bulzacchelli2006,Pisati2020,Yousry2021} to compensate the signal attenuation on the channel at extremely high frequencies, as well as PAM4 \cite{Pisati2020,Yousry2021} which utilizes four voltage levels for signaling on its physical layer, in addition to extremely aggressive technology scaling. The equalization techniques are essential, especially when serial interfaces are designed for lossy channels (e.g., long channel applications), which deteriorate the transmitted signal quality. 

One of the key drivers for serial link applications is the HPC domain. Nowadays, HPC systems are built with relatively small dies (so-called chiplets) integrated within a package to achieve both higher performance and fabrication yield. As described in \cite{amd_chiplet}, short-range links which achieve over Gbps/pin are responsible for the inter-chip data transfer. Data-rates of 112Gbps are available in commercial IP offerings \cite{synopsys_ip}, reflecting on the technology trend where such short-range and high-data rate serial links have become an essential building block for high-performance computing systems.

Besides the bandwidth evolution, energy efficiency has been carefully optimized because I/O power can erode an entire system's energy-efficiency as described in \cite{Poulton2013,choi2018}. For example, the short-range link in \cite{Poulton2019} records 1.17 pJ/bit efficiency at the 25Gbps/pin data rate. Unlike long-reach serial links such as \cite{Pisati2020}, this interface leverage short channel traces for low signal attenuation (up to -8.5dB), allowing for achieving robust operations with simple circuitry. Moreover, at a high data rate, energy per-bit is low even in the presence of steady current consumed by the I/O circuits.  

Even though energy per bit is low, the level of I/O power required by serial links designed for high performance systems is not acceptable for IoT end-devices because the total chip power has to be bound at a few mW. In fact, even a high-speed interface designed for short traces would dissipate more than tens of mW (e.g., 28mW =1.17pJ/bit*25Gbps \cite{Poulton2019}), which is substantially larger than expected average power consumption for an entire IoT end-device SoC.

More closely related to our application target, the work in \cite{choi2015,choi2018} reports a low-power serial interface that achieves extremely low energy consumption while maintaining a few Gbps of data rate. This interface's main feature is to utilize near-threshold voltages (around 0.5V), which are widely adopted in digital circuits and have succeeded in improving energy efficiency by tenfold compared to conventional nominal voltages (e.g., 1.2V) \cite{nearth}. A multiplexed transceiver with a parallel data path connected to a data trace and activated at different clock phases compensates for the degraded gate delay at the near-threshold voltage level. If a transceiver has eight paths, the clock frequency is 1/8 for the same data-rate target. A PLL for the near-threshold voltage operation is also implemented.  As a result, the link can operate at 0.45V and achieves 0.29pJ/bit of the efficiency at 1.2Gbps with a power consumption of 0.348mW. This is a promising example of a low-power serial link that achieves the very high energy efficiency required by IoT end-nodes. However, this link is demonstrated in a stand-alone test chip and has not been integrated in an SoC. 

Another important concern to achieve an efficient link is how the clocking scheme is implemented. If two chips communicating with each other have independent clocks, there is no guarantee that the transmitter's data synchronize to the clock domain on the receiver side. One of the well-known ways to manage this issue is to implement a clock data recovery (CDR) scheme on the receiver side to extract the clock phase information of the transmitted data at the cost of extra circuit overheads\cite{razavi2002}. Another method is to forward the transmitter clock to the receiver, then manipulate its delay to achieve appropriate phases. According to \cite{Poulton2013}, clock forwarding can realize a simpler mechanism than a CDR, resulting in less power/area overheads. Indeed, the low-power link mentioned above \cite{choi2015,choi2018} also utilizes this idea.

Regardless of the continuous research efforts on serial interfaces, most of these systems reported in the literature are just stand-alone and not integrated into real microcontrollers. Indeed, it is common to implement a link system for proof of concept, which requires several additional voltage sources \cite{choi2018} and a test code generator for the transmitter \cite{choi2018,Poulton2019,Pisati2020}. The link in \cite{choi2018} requires an external reference voltage to determine the output swing's amplitude. Integrating an extra voltage source into a chip imposes area/power overheads ignored in that work.  Also, in actual system integration, a transceiver must cooperate with on-chip memory storing the transferred data and a CPU unit issuing the data transaction. However, the links directly connected to a test code generator do not implement additional modules corresponding to such essential interface functionality.  Moreover, the stand-alone system test does not offer any insight into optimization opportunities by exploiting existing support circuits in the rest of the chip and considering the limitations critical for edge-node devices. Even if it is believed that the clock forwarding systems can achieve lower-power consumption than CDRs as previously described, the additional pads needed for clock transfer erode the small pin resources of microcontrollers.

To address the issues mentioned above, we propose a low-power serial link interface for IoT end-nodes, which requires no additional power source and clock forwarding pins, and we integrate it into an open-source microcontroller \cite{pulpissimo}. Unlike state-of-the-art interfaces in the literature, we implement  all the required control modules for our link, which can be managed via an on-chip CPU and allow communicating to an on-chip memory across the link. This means our design is a complete chip-to-chip communication link, including all the hardware and software layers. The link utilizes available resources in the SoC to reduce area overhead, and most of the digital parts exploit a low clock frequency, one-fourth of the data frequency, to reduce power overheads. 

It is important to mention that, although there are some reports of low-power SoCs including serial interfaces \cite{wright2020,Amid2020}, they do not reveal the architectural part of the interface itself and do not present any evaluation of the performance of the interface alone, which is the main focus of our work. 

\section{System Overview}\label{sec:overview}

\begin{figure}[t]
\centerline{\includegraphics[width=\linewidth]{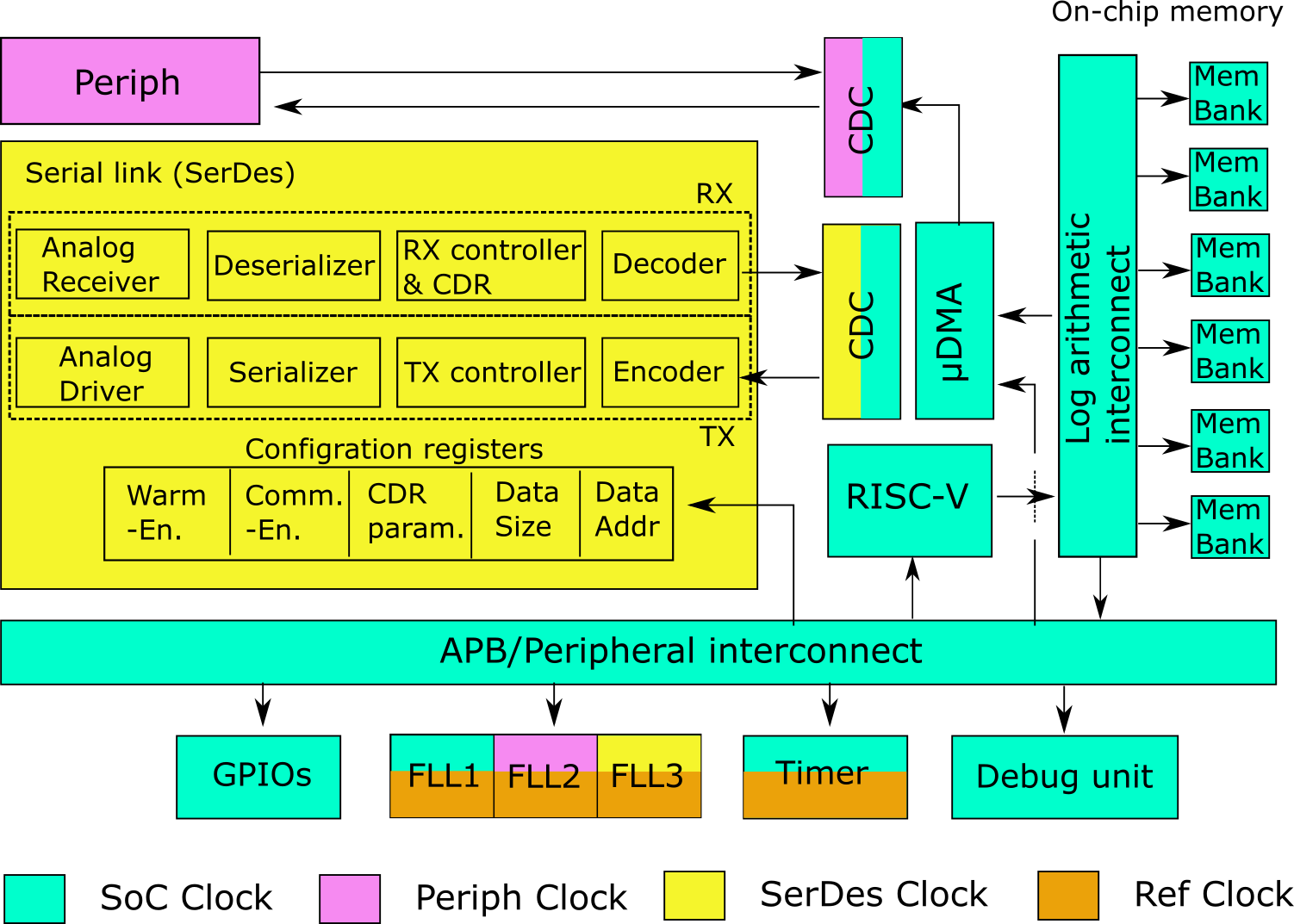}}
\caption{High level architectural block diagram of the system overview}
\label{socarc}
\end{figure}

Fig. \ref{socarc} shows an overview of the SoC hosting the proposed serial interface. It is built around a RISC-V core optimized for energy-efficient digital signal processing served by a word-level interleaved L2 memory and an autonomous input/output subsystem ($\mu$DMA) \cite{udma}. Three frequency locked loops (FLLs) generate the internal clocks. The SoC includes other basic modules: a timer, a debug unit, UART (Periph in Fig. \ref{socarc}), and programmable GPIOs. A JTAG interface (omitted from the figure) accesses the debug unit. It can boot application data into the on-chip memory and start executing the loaded program with the RISC-V core. The SoC has three main clock-domains, namely ``{\em SoC Clock,}'' ``{\em Periph Clock,}'' and ``{\em SerDes Clock}.'' Each domain is driven by a dedicated FLL multiplying the given reference clock. The clock domain crossing (CDC) modules implemented here are asynchronous FIFOs.

The on-chip memory is composed of four 32-kB word-level interleaved shared banks and two 16-kB private banks. The interleaved banks allow four times higher bandwidth with respect to a traditional single/port architecture. Thus, this memory feature is suitable for managing access requests by a high-speed serial link interface and other functional units (e.g., CPU). Also, as the RISC-V core does not have an instruction cache, the private banks are dedicated to the core. This strategy avoids instruction access to conflict against data access by other modules.

The $\mu$DMA provides high-speed autonomous data transfers between the on-chip memory and peripherals. It has two 32-bit ports directly connected to the on-chip memory via the logarithmic interconnect, one dedicated to the write channel, and one to the read channel being the two completely decoupled from each other, delivering 1.6 Gbps at 50 MHz each. The RISC-V core is responsible for programming logical channels, implemented through APB mapped control and status registers available for every peripheral connected to the $\mu$DMA (e.g., I2S, I2C, SPI). When the logical channels are enabled, the physical channels automatically route data to/from the on-chip memory through the read and write physical channels which are shared among all the peripherals. 

The SerDes, integrated into the $\mu$DMA just like any other peripheral is composed of the transmitter (TX), the receiver (RX), and configuration registers mapped on the advanced peripheral bus (APB) used to store enable signals and the address/size of the communicated data. As shown in the block diagram, the clocking scheme is based on a CDR to save the chip's limited pin resources. It is essential to mention that the dedicated FLL for ``{\em SerDes Clock}'' is implemented because of testing purposes. We can omit it when the output of ``{\em Periph Clock}'' can be shared with the SerDes interface. Also, note that the SerDes data rate should always be lower than the one of the $\mu$DMA. Otherwise, the DMA’s effective-bandwidth is limited while we have the SerDes’s power overhead operating at unneeded bandwidth.

\section{SerDes interface}\label{sec:ser_des}
\begin{figure}[t]
\centerline{\includegraphics[width=\linewidth]{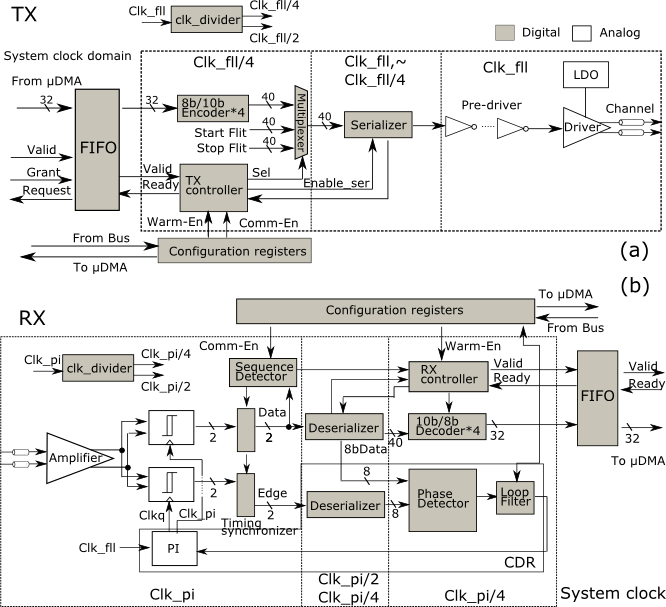}}
\caption{Architectural block diagram of the SerDes interface (a)TX (b)RX}
\label{link_arc}
\end{figure}

%\subsection{Entire architecture}
Fig. \ref{link_arc} shows a detailed block diagram of the SerDes. The TX is composed of the 8b/10b encoders  \cite{8b10b}, TX controller, 40:1 serializer, pre-driver, and the driver. The RX consists of an analog amplifier, analog comparators, timing synchronizers, deserializers, an RX controller, a CDR circuit, and the 10b/8b decoders\cite{8b10b}.  The differential signaling is used for the data channel. This signaling brings the design simplicity for the analog drivers and robustness against common-mode noise. Single-ended signaling with the ground reference voltage imposes additional design complexity and implementation footprint as reported in \cite{Poulton2013}. Since we do not employ a clock forwarding scheme, four analog pads (two for the RX and two for the TX) are necessary for the physical data channel. Also, as the target is miniaturized IoT edge nodes, expected PCB traces would not be longer than a few cm (e.g., less than 5cm), and the operating frequency is not as high as other high-performance applications (less than a few Gbps). So channel equalization is not needed, as verified later by back-annotated simulations in Section \ref{sec:measure}. 

It is worth noting that, by utilizing the ``{\em Periph Clock}'' domain, the SerDes does not require a dedicated clock source.  This clock sharing strategy contributes to avoiding the power and area overhead imposed by an additional clock generation circuit.

\subsection{TX design}
At the TX, 40-bit ``{\em Start flit,}'' ``{\em Stop flit,}'' and the main body of the communication are serialized and transmitted. The ``{\em Start flit}'' and ``{\em Stop flit}'' notify the transferred data’s start and endpoint, respectively. A multiplexer, pictured in the left part of Fig. \ref{link_arc} (a) before the serializer, selects one of the 40-bit flits and sends it to the serializer. The four parallel 8b/10b encoders ensure that the serialized data is DC-balanced and its disparity is less than $\pm 2$. The TX controller is a finite state machine that manages these functionalities’ timing according to the FIFO handshaking signals from the interface between the SerDes and the $\mu$DMA and the enable signals from the configuration registers.  The TX clock is provided by the FLL and divided by two and then four.  The encoders, multiplexer, and controller utilize ``{\em Clk$\_$fll/4}'' to reduce the power consumption. 

\begin{figure}[t]
\centerline{\includegraphics[width=0.8\linewidth]{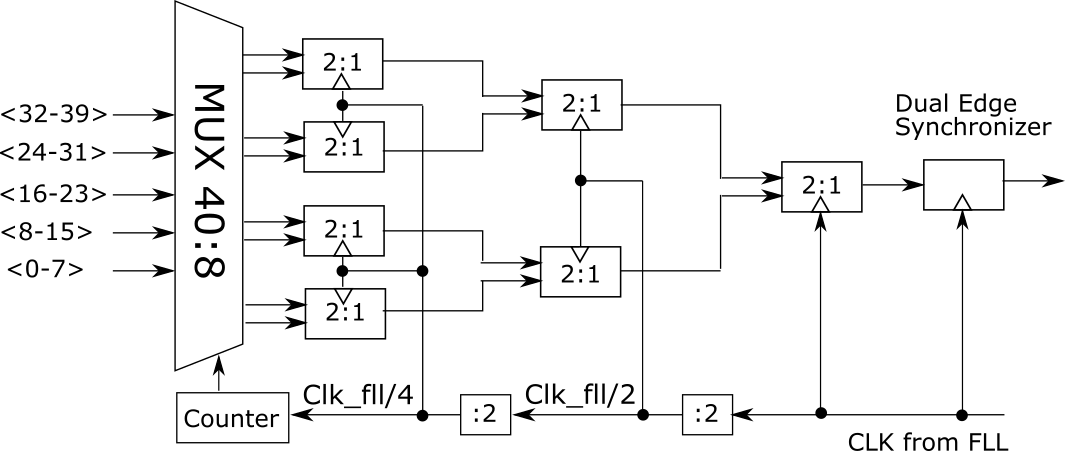}}
\caption{Architectural block diagram of the serializer}
\label{serializer}
\end{figure}

Fig. \ref{serializer} depicts the detailed architecture of the 40:1 serializer \cite{dazzi2018}. According to the counter output synchronizing to ``{\em Clk$\_$fll/4}'', the multiplexer selects a set of 8-bits over the 40-bit word. Then, the tree of 2:1 modules serializes the selected one. The serialized data synchronize at both the falling and the rising edge of ``{\em Clk$\_$fll}'' (DDR). 

The output of the counter informs the TX controller about the progress of the serialization. For example, when the counter's value is 0, the first 8-bits ($<0-7>$) is serialized. When it is 4, the serialization is completed for a 40-bit flit.  

\begin{figure}[t]
\centerline{\includegraphics[width=0.75\linewidth]{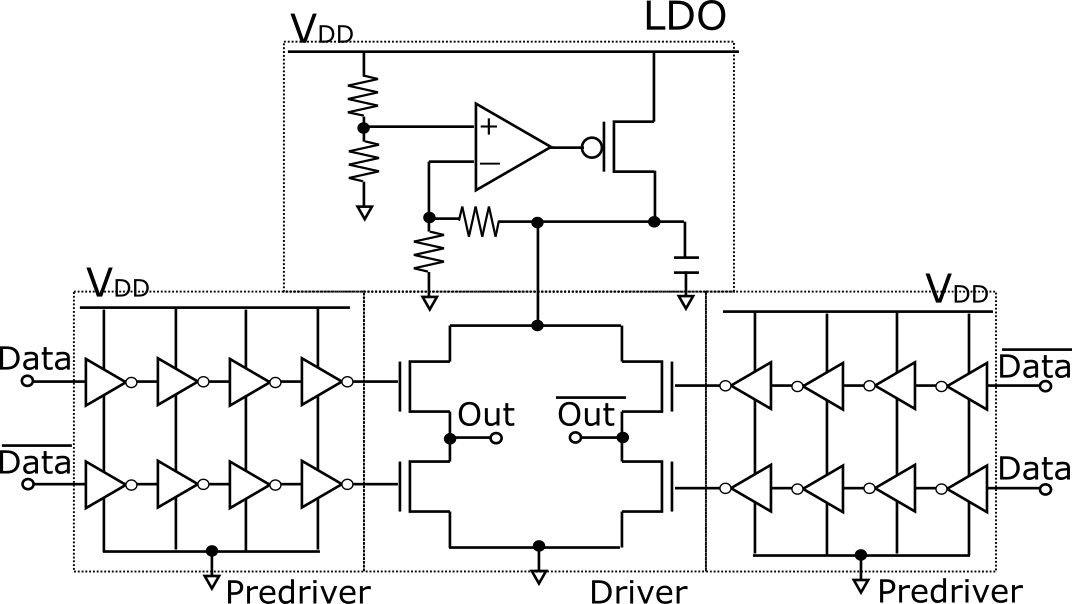}}
\caption{Circuit diagram of the TX driver and predriver}
\label{driver}
\end{figure}

The driver is composed of a voltage-mode differential driver with a chain of inverters acting as a pre-driver\cite{dazzi2018} (Fig. \ref{driver}). It is powered by a low-dropout regulator (LDO) to operate at a lower voltage swing than the full voltage supply of 1.2V. The transceiver is source-terminated (by adequately sizing the transistors in the driver) to match a differential impedance of 100$\Omega$.

It is important to mention that this work's design approach is different from implementing a power-hungry LVDS \cite{lvds} compliant peripheral driving 3.5mA of current for each line. Thanks to the design strategies such as using a high impedance RX input (as opposed to 100$\Omega$ differential) and setting the driver MOS’s output resistance to 50$\Omega$ (instead of adding series resistors), the driver implemented here consumes much less power, as shown in section \ref{sec:measure}. 

\begin{figure}[t]
\centerline{\includegraphics[width=0.75\linewidth]{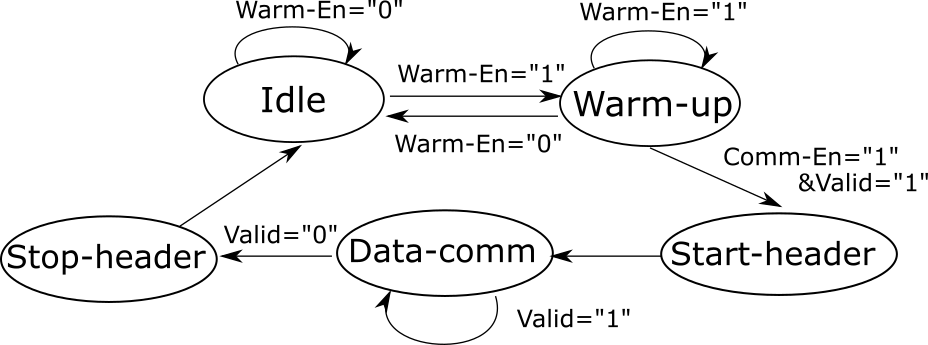}}
\caption{Finite state machine of the TX controller}
\label{txcontroller}
\end{figure}

Fig. \ref{txcontroller} shows the finite state machine of the TX controller.  Firstly, the TX is in the idle state. Then, by asserting ``{\em Warm-En,}'' the state is changed to ``{\em Warm-up},''  where the TX outputs a DC-balanced training sequence including its clock phase information (warm-up mode). According to the sequence, the CDR loop manipulates the RX clock phase to synchronize to the TX data before establishing the communication.  In case ``{\em Comm-En}'' and ``{\em Valid}'' signals are ``1'', the state is in ``{\em Start-header},'' which sends the ``{\em Start flit}.'' After the header transfer, the state automatically goes to ``{\em Data-comm}'' (data-comm mode), then the main part of the data communication is started. The multiplexer’s output is updated every 20 cycles during this mode as the serial data synchronizes at DDR. This control is realized by checking the counter value from the serializer and activating the encoder at the proper timing. When the ``{\em Valid}'' signal is negated, the ``{\em Stop flit}'' is sent. Finally, the state is back to the idle one.

\subsection{RX design}
At the RX, the input signal is sent to the amplifier first. Then it is captured by the comparators, which restore the even and odd bit data from the channel (Fig. \ref{link_arc}(b)). After the deserialization, the 32-bit data is decoded, and the original 40-bit data is transferred to the $\mu$DMA through the asynchronous FIFO interface. Since the data communication begins from ``{\em Start flit}'' and ends at ``{\em Stop flit},'' the sequence detector monitors whether they arrive at or not. This mechanism is realized by checking 11011111 ($K_{27,7}$ in \cite{8b10b}) for ``{\em Start flit}'' and 10111111 ($K_{29,7}$) for ``{\em Stop flit.}'' According to the detector's information, the RX controller manages the deserializer and 10b/8b decoders for the transferred data's main body. The decoded data is sent to the FIFO with the ``{\em Valid}'' signal according to the ``{\em Ready}'' signal from it.

The CDR scheme is composed of the phase detector, loop filter, and phase interpolator (PI) adjusting the FLL clock phase. The phase detector compares ``{\em Data}'' captured by the normal clock (``{\em Clk\_pi}'') with ``{\em Edge}'' synchronizing to a quadrature clock (``{\em Clkq}'') \cite{razavi2002}. This comparison produces "{\em Early}" and "{\em Late}" signals. They are accumulated and then used to control the PI.  For the other modules in the RX, the generated clock is divided by four (``{\em Clk$\_$pi/4}'') and two (``{\em Clk$\_$pi/2}'').  As shown in Fig. \ref{link_arc}(b), The RX controller, decoders, and some parts of the CDR loop synchronize to ``{\em Clk$\_$pi/4}'' to reduce the power consumption.

\begin{figure}[t]
\centerline{\includegraphics[width=\linewidth]{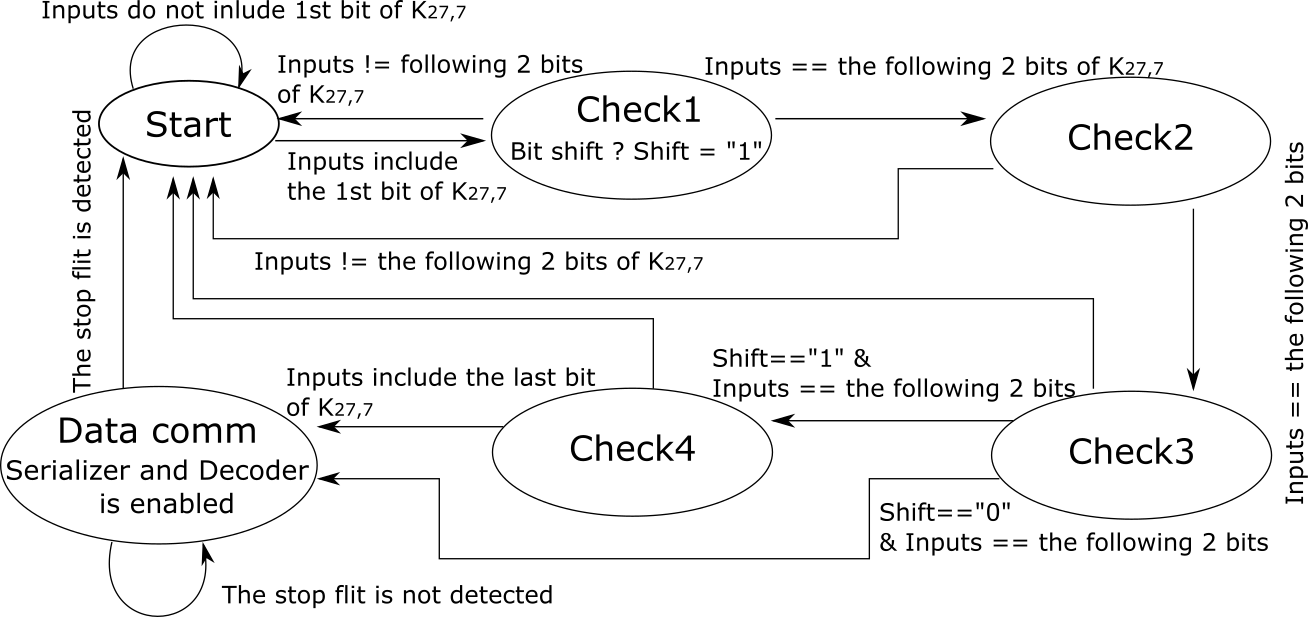}}
\caption{State machine of the sequence detector}
\label{sequence_detector}
\end{figure}

%\subsubsection{Sequence detector}
The sequence detector checks the even and odd bits captured by the analog comparators to activate the entire RX when the ``{\em Start flit}'' arrives. The sequence detector is composed of a finite state machine, as shown in Fig. \ref{sequence_detector}. 

The state of the detector is changed when $K_{27,7}$ arrives. In other words, when 11 out of 11011111 is detected, the next state is ``{\em Check1}.'' After this, if the following two bits are 01, the state is updated to ``{\em Check2}.'' When the full 8-bit of $K_{27,7}$ is detected, the state goes to ``{\em Data-comm}.'' The RX controller enables the deserializer and decoder at this state. Also, during the data communication, it is monitored whether the ``{\em Stop flit}'' arrives or not with a similar procedure. When it is detected, the state of the detector goes back to ``{\em Start}.''

It is important to mention that the RX has to consider the possibility of a data shift; even if the TX sends an even bit, the RX might capture it as an odd-bit. For example, the sequence of 11011111 might be captured as x1 10 11 11 1x. The state machine holds the bit shift information as the signal ``{\em Shift}'' to manage such a situation.  The ``{\em Check4}'' state checks additional 2 bits of the input when the shift occurs.

\begin{figure}[t]
\centerline{\includegraphics[width=0.65\linewidth]{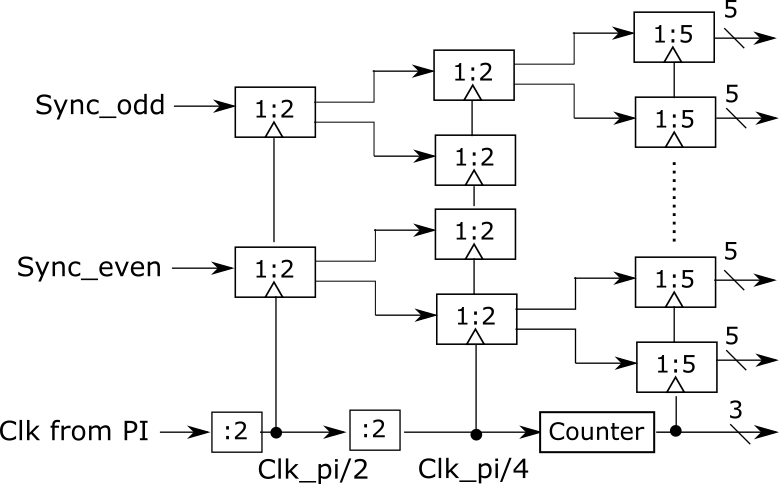}}
\caption{Architectural diagram of the deserializer}
\label{deser}
\end{figure}

%\subsubsection{Deserializer and RX controller}
Fig. \ref{deser} depicts the adopted deserializer architecture.  Firstly, the tree of 1:2 modules restores 8-bit data, then, 1:5 modules complete the 8:40 deserialization. The counter selects one bit of the 1:5 modules' output. So, as similar to the TX, the counter value indicates deserialization progress utilized by the RX controller. The 8-bit output is forwarded to the CDR's phase detector together with the 1:8 deserialization of the edge samples \cite{dazzi2018}. Note that, unlike digital circuits, it is difficult to define analog macros' timing information for automatic P\&R tools, possibly resulting in long paths and timing violations.  Hence, in Fig. \ref{link_arc}, the modules named ``Timing synchronizers'' are placed between the comparators and the deserializers. They also adjust the bit shift according to the ``{\em Shift}'' signal from the sequence detector. Hence the deserializer always receives the even and odd bit correctly. 

Before the data communication, the controller activates only the CDR loop components by asserting ``{\em Warm-En}'' (warm-up mode). After the loop settles, the configuration registers assert the enable signal for the sequence detector (``{\em Comm-En}'').  When the ``{\em Start flit}'' arrives, the controller state is in the data-comm mode enabling the entire deserializer. The decoders update their output when 40-bit data is ready. As mentioned before, the counter in the deserializer indicates this timing. The ``{\em Valid}'' signal is also generated after the latency of the decoders. When the ``{\em Stop flit}'' arrives, the controller disables the 8:40 deserializer and decoders if ``{\em Warm-En}'' is still asserted. In case that all the enable signals for the RX are negated, the RX is in idle mode. 

%\subsubsection{Amplifier and Comparator}
The analog front-end employs a differential amplifier that feeds the data and edge comparators. 
The comparator is based on the double-tail comparator described in \cite{comparator}. The advantage behind the double-tail comparator is the reduction of transistor stacking, allowing low supply voltages. 

\begin{figure}[t]
\centerline{\includegraphics[width=0.9\linewidth]{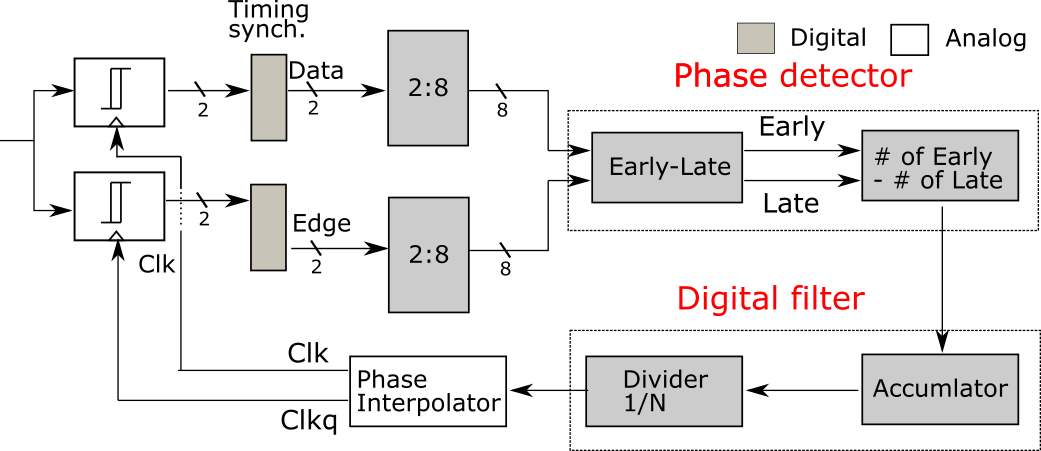}}
\caption{Architectural diagram of the CDR loop}
\label{CDR}
\end{figure}

%\subsubsection{Clock Data Recovery module}
Fig. \ref{CDR} shows an architectural diagram of the CDR loop. First, the ``Early-Late'' part compares the 8-bit ``{\em Data}'' forwarded by the 2:40 deserializer with the 8-bit ``{\em Edge}'' generated by the dedicated 2:8 module. The ``Early-Late'' is composed of seven parallel Alexander phase detectors \cite{alexander1975}. Then, the number of ``{\em Early}'' is subtracted by the number of ``{\em Late}.'' The result is accumulated and divided by 1/N (N=1,2,4,8,..., 128). According to the result, the PI shifts the clock phase for both ``{\em Clk}'' and ``{\em Clkq}.'' We define the resolution of this adjustment as $2\pi/32$ in the current design. The PI is a charge-based interpolator based on \cite{elnaqib2020}. Since the TX clock generation is in another chip, the actual frequency slightly varies from the RX one. However, this is detected as a phase difference continuously increasing over time corrected by the CDR loop. In this respect summing the Early/Late signals obtained from the deserialized data and edge sampling is advantageous compared to voting since the CDR loop’s bandwidth is improved, and larger differences between the TX and RX FLLs can be tracked.

\section{Software model for the communication}\label{sec:chips_sync}
\subsection{Software control for the SerDes}
As previously described, the SerDes operates in three modes: idle, warm-up, and data-comm.   The enable signals (``{\em Comm-En}'' and ``{\em Warm-En}'') selects these three modes. The signals are stored in the memory-mapped configuration registers by the application running on the core. 
Before starting a $\mu$DMA transaction, the core sets the SerDes’ configuration registers (Table \ref{config_reg}), modifying the address pointer in the on-chip memory and the target data’s size to be sent. Then, the SerDes is set to the warm-up state by the ``{\em Warm-En}'' signal. 
After the CDR loop is ready, ``{\em Comm-En}'' is asserted by the application, and the $\mu$DMA automatically sends the data to the SerDes according to the address/size information. The TX module transmits the data from the $\mu$DMA to another chip.  The RX module in another chip captures the transferred data and delivers it to the $\mu$DMA. According to the configuration registers' information, the $\mu$DMA sends the received data to the RX buffer in the on-chip memory.

\begin{table}[t!]
	\centering
	\caption{Configuration registers}
	\label{config_reg}
	\begin{tabular}{|c|c|}
		\hline 
		Reg name & Description \\ \hline
		TX Data Addr & The start address of the transferred data  \\ \hline 
		RX Data Addr & The start address of the RX buffer region \\ \hline
		TX Data Size & The data size to be transferred \\ \hline
		RX Data Size & The data size to be stored \\ \hline
		Warm-En & The warm-up enable signals \\ \hline
		Comm.-En & The communication enable signals \\ \hline
		CDR param.(N in Fig.\ref{CDR}) & The CDR parameters \\ \hline
	\end{tabular}
\end{table}

\subsection{Chip-to-Chip Communication}
\begin{figure}[t]
\centerline{\includegraphics[width=\linewidth]{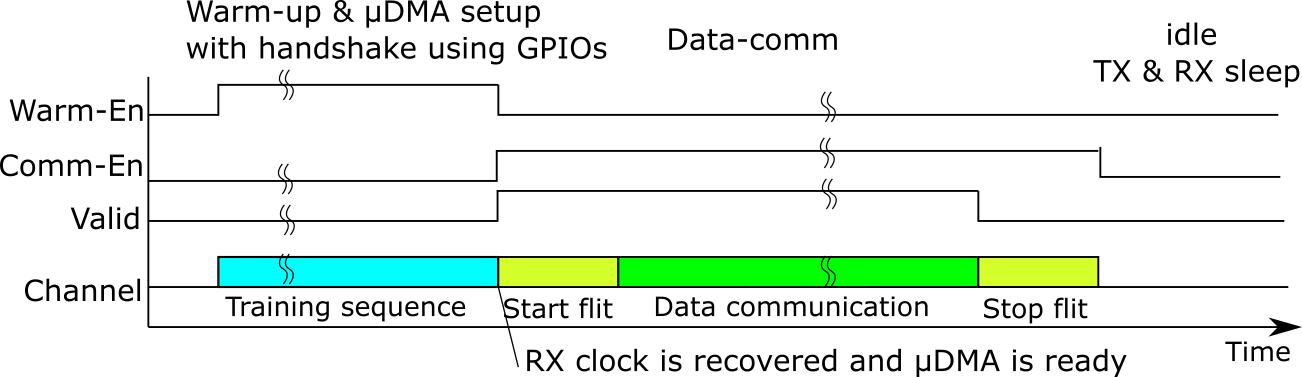}}
\caption{Timing chart of the data transfer}
\label{timing-tx}
\end{figure}

In actual systems, the TX and RX should be activated only when the data communication occurs because of its non-negligible power overhead. Fig. \ref{timing-tx} shows a conceptual timing chart for the data communication. After the data transaction, the TX and RX turn to sleep.

The problem lying here is that while the SerDes is in idle mode, the CDR loop does not track the TX data. So the channel has to be warmed up again when the next data communication occurs. Also, before the SerDes communication, the $\mu$DMA module must be set up for its data transaction from/to the on-chip memory. In this scenario, the TX has to know whether the RX in another chip is ready. We solve this concern with GPIOs, which are generally available on microcontrollers. Here we assume that the TX(RX) chip controls ``{\em GPIO0}''(``{\em GPIO1}'').

\begin{algorithm}[t]                      
\caption{Programming model for the case that the TX invokes the communication}
\label{alg1}                          
    \begin{algorithmic}[1] 
     \STATE $\sim\sim\sim$ TX Chip $\sim\sim\sim$
        \STATE Setup the direction of {\em GPIO0}.
        \STATE Prepare the transferred data in the on-chip memory.
        \STATE Setup $\mu$DMA (Write configuration registers for data size \& address.)
        \STATE {\em Warm-En} $\Leftarrow 1 $
        \STATE {\em GPIO0} $\Leftarrow 1 $
        \WHILE{{\em GPIO1} $!= 1$}
                \STATE Wait.
        \ENDWHILE
        \STATE {\em Comm-En} $\Leftarrow 1 $
    \STATE  $\sim\sim\sim$ RX Chip $\sim\sim\sim$
        \STATE Setup the direction of {\em GPIO1}.
        \STATE Interrupt happened by {\em GPIO0}.
        \STATE Prepare an RX buffer region in the on-chip memory.
        \STATE Setup $\mu$DMA (Write configuration registers for data size \& address.)
        \STATE {\em Warm-En} $\Leftarrow 1 $
        \WHILE{The RX clock is not ready}
            \STATE Wait.
        \ENDWHILE
        \STATE {\em Comm-En} $\Leftarrow 1 $
        \STATE {\em GPIO1} $\Leftarrow 1 $
    \end{algorithmic}
\end{algorithm}

The first situation that has to be considered is when a TX chip (e.g., a sensor module) invokes the data communication to an RX chip (e.g., a microcontroller).  
Algorithms \ref{alg1} depicts a programming model to synchronize the two chips in this case. The lines from 1 to 10 show the pseudo-code for the TX chip.  Also, the codes for the RX chip are noted from 11 to 21.
In the TX chip, the ``{\em Warm-En}'' signal is asserted after preparing the transferred data and setting $\mu$DMA (Lines 2$\sim$5), and the training sequence is sent to the RX chip. Then ``{\em GPIO0}'' notifies the RX that the TX chip is ready for the data communication (Line 6). In the RX chip, ``{\em GPIO0}'' from the TX causes interruption (Line 13) and then sets up the link communication.  After asserting RX's ``{\em Warm-En,}'' the SerDes has to wait until the RX clock recovers, as shown in the lines from 17 to 19. Once everything for the communication is ready, ``{\em GPIO1}'' is asserted to notify the TX chip that the RX can accept the data transfer. With this ``{\em GPIO1}'' signal, the TX starts to transmit the data by asserting the ``{\em Comm-En}'' signal. 

We also have to consider the case when an RX chip (e.g., a microcontroller) requests the data communication from a TX chip (e.g., an off-chip memory) as shown in Algorithm \ref{alg2}. In this scenario, the RX chip is set up first, and then it tells the TX via ``{\em GPIO1}'' that the data communication takes place (Lines 1$\sim$6). The TX chip starts to set up $\mu$DMA and the configuration registers by receiving the RX interrupt signal, and it sends the training sequence (Lines 18 and 19). Also, the TX chip notifies the RX of the start point of the warm-up mode (Line 20). After that, the RX conducts the clock recovery (Lines 10$\sim$12). When the RX is ready to accept the main data, ``{\em GPIO0}'' is negated (Line 14) to notify the TX that it can start the communication (Line 24). 

\begin{algorithm}[t]                      
\caption{Programming model for the case that the RX invokes the communication}
\label{alg2}                          
    \begin{algorithmic}[1] 
     \STATE $\sim\sim\sim$ RX Chip $\sim\sim\sim$
         \STATE Setup the direction of {\em GPIO1}.
         \STATE Prepare an RX buffer region in the on-chip memory.
         \STATE Setup $\mu$DMA (Write configuration registers for data size \& address.)
         \STATE {\em Warm-En} $\Leftarrow 1 $
         \STATE {\em GPIO1} $\Leftarrow 1 $
         \WHILE{{\em GPIO0} != 1}
            \STATE Wait.
        \ENDWHILE
        \WHILE{The RX clock is not ready}
            \STATE Wait.
        \ENDWHILE
        \STATE {\em Comm-En} $\Leftarrow 1 $
        \STATE {\em GPIO0}  $\Leftarrow 0 $
     \STATE $\sim\sim\sim$ TX Chip $\sim\sim\sim$  
         \STATE Setup the direction of {\em GPIO0}.
         \STATE Interrupt happened by {\em GPIO1}.
         \STATE Setup $\mu$DMA (Write configuration registers for data size \& address.) 
         \STATE {\em Warm-En} $\Leftarrow 1 $
         \STATE {\em GPIO0} $\Leftarrow 1 $
         \WHILE{{\em GPIO0} == 1 \& {\em GPIO1} ==1}
            \STATE Wait.
        \ENDWHILE 
     \STATE {\em Comm-En} $\Leftarrow 1 $
    \end{algorithmic}
\end{algorithm}

\section{Chip implementation} \label{sec:chip}
\begin{figure}[t]
\centerline{\includegraphics[width=0.7\linewidth]{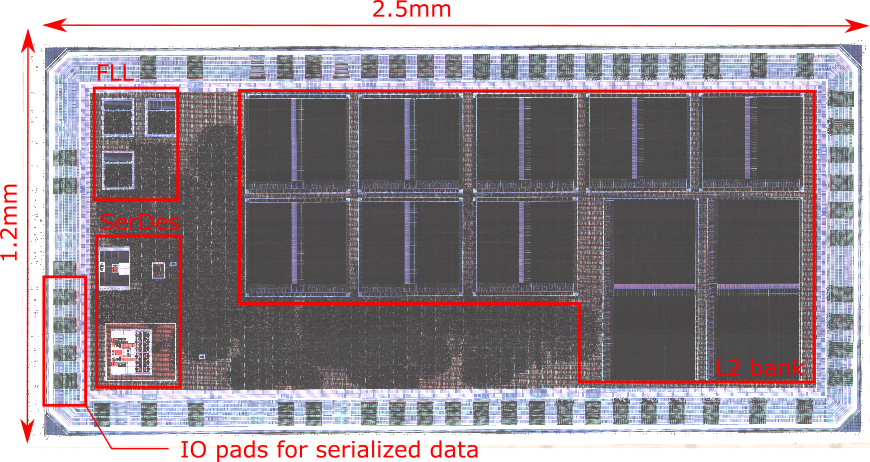}}
\caption{Chip photograph of the whole system}
\label{chip_photo}
\end{figure}

We implemented and fabricated the above-mentioned whole SoC, including the SerDes in 65-nm CMOS technology (UMC 65-nm) \cite{umc65}. The chip photograph is shown in Fig. \ref{chip_photo}. The analog signals are connected to four library I/O cells featuring a built-in 50-ohm resistor. Two of them are for the RX, and the others are for the TX. Synopsys Design Compiler 2018.06-SP1 and Cadence Innovous v15.20 were employed for the synthesis and P\&R. The used chip package is QFN56. 

\begin{figure}[t]
\centerline{\includegraphics[width=0.55\linewidth]{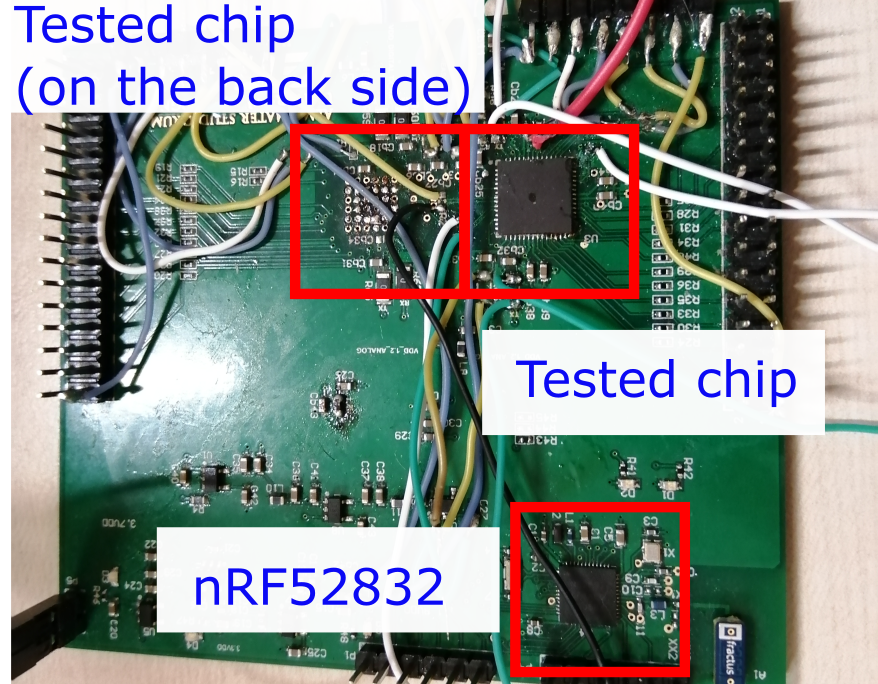}}
\caption{Photograph of the evaluation board}
\label{board_photo}
\end{figure}

The nominal voltage and operational frequency of the SerDes are 1.2V and 400MHz, respectively. Hence, the current design's target bandwidth is 0.8 Gbps as the data transfer is performed at DDR. The nominal voltage is used for both digital and analog circuits because, as previously described, adding extra voltage sources increases system costs which should be avoided for embedded microcontrollers.  The default frequency of the microcontroller is 50MHz. So the data rate of the SerDes is lower than one of the $\mu$DMA. Hence the $\mu$DMA can fully saturate the bandwidth available at the data link.

According to a Matlab simulation, the minimum number of N (divider ratio) in Fig. \ref{CDR} for the loop convergence is equal to four at the nominal frequency target. So, in our test, we set N=4. Also, in our system-level simulation, the maximum frequency deviation between RX and TX that the CDR loop can track is approximately 0.4\% with this setup.

We implemented a test board with the two fabricated chips.  The board also has a Nordic Semiconductor nRF52832 SoC (Fig. \ref{board_photo}). The nRF52832 chip is responsible for sending a JTAG compliant sequence to the test chips to load application data into the on-chip memory. Once the boot ends, the JTAG connection activates the core in the test chip, and a test code for the SerDes runs. The TX and RX channel trace between the two chips is kept at 2cm as the link targets small IoT end-node boards. For flexible measurements, an external function generator (Agilent 33220A) provides the reference clock for the FLL. Nevertheless, the board allows mounting a crystal oscillator and utilizing it. The power supply pins for the SerDes are connected to wires to measure power consumption.

\section{Chip measurement results and Efficiency evaluation}\label{sec:measure}
\subsection{Measurement result}
\begin{figure}[t]
\centerline{\includegraphics[width=0.8\linewidth]{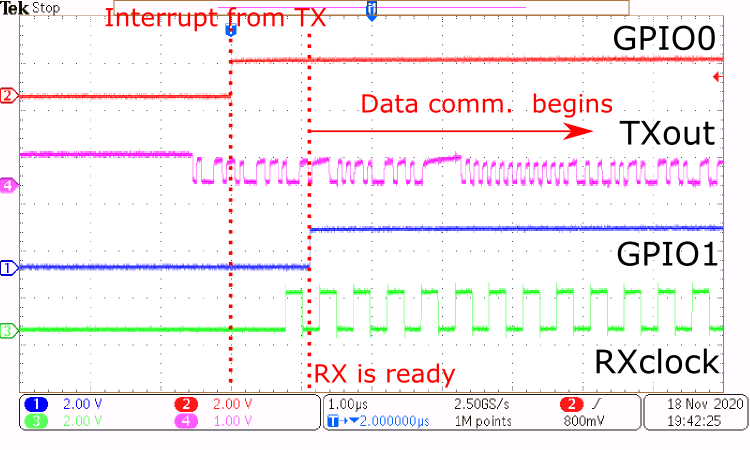}}
\caption{Waveform of the two chip synchronization}
\label{waveform_sync}
\end{figure}

Fig. \ref{waveform_sync} shows the waveform of the two chip communication synchronization described in Algorithm \ref{alg1}. To improve the readability of the figure we lowered the SerDes' clock frequency to 8MHz, and we disabled the CDR loop (i.e., only the PI is active). Also, the frequency of the monitored PI output is divided by four. As we can see, the TX sends the interrupt signal through ``{\em GPIO0}'' first, and then the RX starts activating the RX clock (i.e., the PI begins to output the clock signal). When the RX is ready to accept the data from TX, ``{\em GPIO1}'' is asserted, and the TX starts to send the data sequence at a low-voltage swing (0.44V here).

\begin{figure}[t]
\centerline{\includegraphics[width=\linewidth]{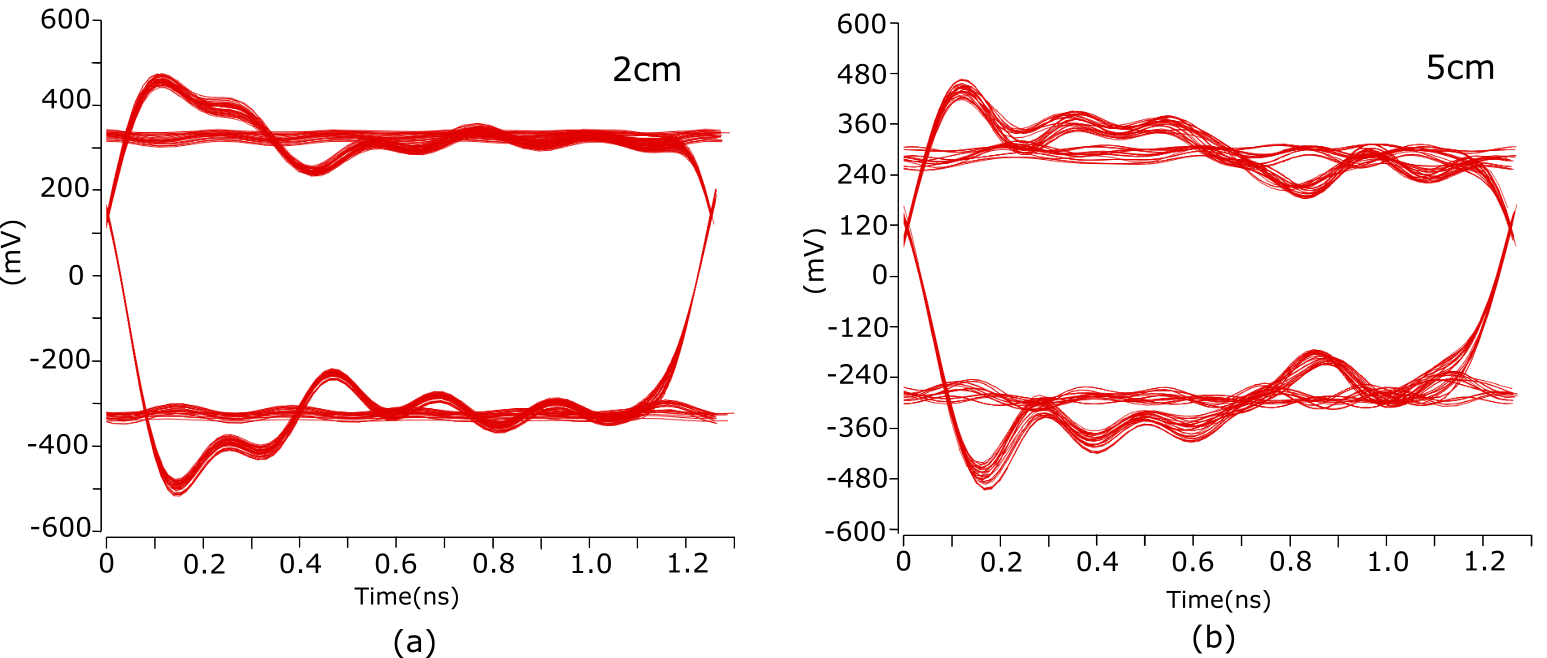}}
\caption{Simulated eye-diagram (0.8Gbps, the RX input impedance is high):(a) with the 2cm trace (b) with the 5cm trace}
\label{eyediagram}
\end{figure}

As we do not have an equalization mechanism in our SerDes topology, it is necessary to verify the TX driver’s signal integrity under PCB trace’s attenuation. We simulate the driver's eye-pattern with a micro-stripe model obtained by the parameter calculator \cite{pcb_calc}. Here we implement the 2cm and 5cm emulation of a PCB differential channel and connect them to the driver. Figs. \ref{eyediagram} (a) and (b) show the simulated eye diagrams depicted with an input sequence generated by the microcontroller. These figures are obtained by repeatedly sampling and superimposing 150 Unit Intervals (UI) of the TX signal. The UI period here is 1.25ns. As can be seen, the longer trace attenuates the signal more. Indeed, the eye height with the 5cm trace is 0.386V, while for 2cm it is 0.418V. Nevertheless, both eyes are open without the need for equalization at this frequency.

Table \ref{power_table} shows the SerDes' measured power consumption at 400MHz of the operational frequency. Since it is difficult to capture the exact timing of the communication mode with a multimeter, we alternatively show a simulation result of the RX digital module obtained by a back-annotated simulation for this operating mode. We checked that the difference between the simulation and the real chip is only 5.5\% in the warm-up mode. According to the table, the total power consumption at the data communication mode is 5.2mW. The estimated energy efficiency at 0.8 Gbps is 6.5pJ/bit.  Also, if we assume that the clock to the digital modules is completely gated at the standby mode and the analog modules are power-gated, the standby power is just 2$\mu$W. Note that LVDS compliant drivers need to provide 3.5mA. Regarding the actual product in \cite{lvds-product}, the power consumption of 4 channel drivers can reach 30mW (i.e., 7.5mW per channel) that exceeds the power budgets of most IoT end-nodes. On the other hand, our TX driver consumes only 0.695mW, which is more suitable for the target applications.

\begin{table}[t!]
	\centering
	\caption{Power consumption of the serial link @ 1.2V, 400MHz (* simulation result)}
	\label{power_table}
	\begin{tabular}{|l|l|l|}
		\hline 
		Power consumption & RX & 3.66mW \\ \cline{2-3}
		(Analog parts)    & TX & 0.695mW \\ \hline 
		Power consumption & RX & 0.591*mW : data-comm mode  \\
		(Digital parts)   &    & 0.368 mW : warm-up mode\\ 
		                  &    & 0.001 mW : standby mode \\ \cline{2-3}
		                  & TX & 0.253 mW : data-comm \& warm-up\\ 
		                  &    & 0.001 mW : standby mode \\ \hline
	\end{tabular}
\end{table}

\begin{figure}[t!]
\centerline{\includegraphics[width=0.7\linewidth]{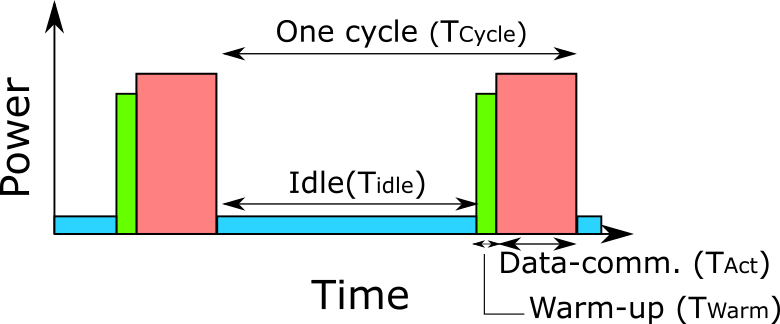}}
\caption{Conceptual timing diagram of the duty-cycled operation}
\label{duty-cycled}
\end{figure}

In case that a required bandwidth is lower than 0.8Gbps, the power consumption is further lowered by utilizing a duty-cycled operation \cite{Anand2014} which periodically turns on the SerDes. Fig. \ref{duty-cycled} shows its conceptual timing diagram. Here, $T_{Cycle}, T_{Act}, T_{Warm}$, and $T_{Idle}$ represent one cycle period, duration of the data-comm, warm-up, and idle mode, respectively. The data communication is conducted until the RX buffer allocated in the on-chip memory is filled up. Then, the SerDes state is back to the idle mode. Here, we also assume that the analog macros' power supply is turned off during the idle state. When it is activated again, the warm-up mode settles the CDR loop with the overhead of $T_{Warm}$. Under these assumptions, $T_{Act}$ is calculated by (Buffer size)/0.8Gbps. Also, $T_{cycle}$ is obtained by considering how many cycles are necessary for one second to achieve a target bandwidth (i.e., $T_{cycle}$=(Buffer size)/(Target BW)).  

Since we cannot precisely measure when the CDR loop converges, let us make the worst-case assumption. Here we consider the phase interpolater shifts 16 steps of the RX clock. The input for the phase detector is updated every four cycles. It goes to the accumulator and 1/N divider (N=4). So the output of the digital filter is updated every 16 cycles. Hence, the estimated worst assumption is $16*16*2.5ns=0.64{\mu}s$. Also, we have to consider the programming latency imposed by Algorithm \ref{alg1}/\ref{alg2}. According to the RTL simulation with 50MHz of the microcontroller's frequency, this takes about 0.75$\mu$s. So the total $T_{Warm}$ is 1.39$\mu$s.

As described in \cite{Usami2011}, gating the power supply imposes an energy overhead when the switch turns on again. According to the back-annotated simulation, this overhead from the switch for the analog modules was 120pJ.

\begin{figure}[t]
	\centering
	\definecolor{color1}{rgb}{0.12156,0.25098,0.47843}% ETH blue
\definecolor{color2}{rgb}{0.50980,0.74509,0.11764}% ETH light green 
\definecolor{color3}{rgb}{0.50100,0.50100,0.50100}% gray
\definecolor{color4}{rgb}{0.862,0.078,0.235} % red
\definecolor{color5}{rgb}{0.83,0.302,0.1}

\begin{tikzpicture}
\begin{axis}[%
width=2.5in,
height=1.35in,
scale only axis,
xmin=0,
xmax=64,
xlabel={Buffer size  (\si[per-mode=symbol]{\kilo\byte})},
ymin=6.5,
ymax=7.5,
ytick={5,5.5,6,...,10},
ylabel={Energy per bit (\si[per-mode=symbol]{\pico\joule\per\bit})},
axis background/.style={fill=white},
%title={$\text Mr. Wolf)}$},
%xmajorgrids,
%xminorgrids,
%ymajorgrids,
%yminorgrids,
legend style={legend cell align=left, align=left},
legend entries={{\tiny BW=50Mbps}, {\tiny 100Mbps}, {\tiny 200Mbps}, {\tiny 400Mbps}, {\tiny 600Mbps}}
]

%\addplot [draw=none, mark=*, mark options={solid, gray,scale=0.25},]
%table[row sep=crcr]{%
%	1	60.0092464564725\\
%	2	58.9733678272748\\
%	4	58.834915322142\\
%	6	58.7197382901612\\
%	8	58.0382615219894\\
%	10	57.8784527510667\\
%	20	57.5928968850709\\
%	30	56.2172675778606\\
%	40	56.1847744602664\\
%	50	54.6008699582679\\
%};

\addplot [name path=link_periodic-50, mark=square*, mark options={solid, color4,scale=0.5}, color=color4, thick]
table[row sep=crcr]{%
64	6.550681839 \\
32	6.56432373 \\
16	6.591147461 \\
8	6.644794922 \\
4	6.752089844 \\
2	6.966679688 \\
1	7.395859375 \\
0.5	8.25421875 \\
};

\addplot [name path=link_periodic-100, mark=star, mark options={solid, color1,scale=0.5}, color=color1, thick]
table[row sep=crcr]{%
64	6.530911865 \\
32	6.54432373 \\
16	6.571147461 \\
8	6.624794922 \\
4	6.732089844 \\
2	6.946679688 \\
1	7.375859375 \\
0.5	8.23421875 \\
};

\addplot [name path=link_periodic-200, mark=*, mark options={solid, color2,scale=0.5}, color=color2, thick]
table[row sep=crcr]{%
64	6.520911865 \\
32	6.53432373 \\
16	6.561147461 \\
8	6.614794922 \\
4	6.722089844 \\
2	6.936679688 \\
1	7.365859375 \\
0.5	8.22421875 \\
};

\addplot [name path=link_periodic-400, mark=triangle*, mark options={solid, color5,scale=0.5}, color=color5, thick]
table[row sep=crcr]{%
64	6.515911865 \\
32	6.52932373 \\
16	6.556147461 \\
8	6.609794922 \\
4	6.717089844 \\
2	6.931679688 \\
1	7.360859375 \\
0.5	8.21921875 \\
};
\addplot [name path=link_periodic-600, mark=otimes*, mark options={solid, red,scale=0.5}, color=red, thick]
table[row sep=crcr]{%
64	6.514245199 \\
32	6.527657064 \\
16	6.554480794 \\
8	6.608128255 \\
4	6.715423177 \\
2	6.930013021 \\
1	7.359192708 \\
0.5	8.217552083 \\
};
\end{axis}
\end{tikzpicture}%
	\caption{Relation ship among Energy, buffer size, and target bandwidth}
	\label{Eperbit_link} 
\end{figure}
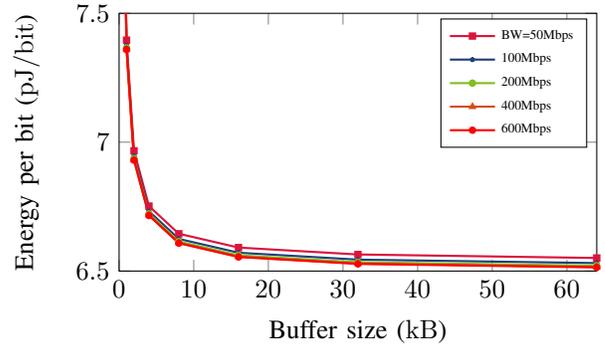

Using the assumptions mentioned above and the values in Table \ref{power_table}, we obtain Fig. \ref{Eperbit_link}. Each of the lines corresponds to the different bandwidth under the duty-cycled operation. As can be seen from the graph, the smaller buffer sizes degrade the energy efficiency. The reason for these results is that idle power dominates the total energy. Also, this graph shows too large buffer sizes do not always contribute to energy efficiency. Indeed, when the buffer size reaches 16KB, the efficiency improvement is almost saturated at all of the bandwidth targets. Hence, 16KB is a good compromise in terms of required memory size and available energy efficiency.

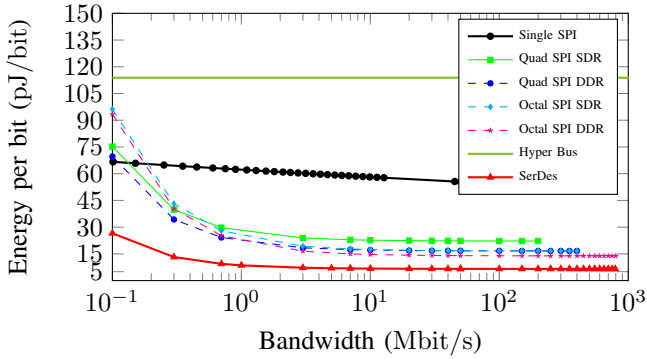
\begin{figure}[t]
	\centering
	\definecolor{color1}{rgb}{0.12156,0.25098,0.47843}% ETH blue
\definecolor{color2}{rgb}{0.50980,0.74509,0.11764}% ETH light green 
\definecolor{color3}{rgb}{0.50100,0.50100,0.50100}% gray
\definecolor{color4}{rgb}{0.162,0.078,0.635}
\definecolor{color5}{rgb}{0.83,0.302,0.1}
\definecolor{color6}{rgb}{1,0.8,0.0}
\begin{tikzpicture}
\begin{axis}[%
width=2.7in,
height=1.4in,
scale only axis,
xmin=0.1,
xmax=1E03,
xmode= log,
xlabel={Bandwidth (\si[per-mode=symbol]{\mega\bit\per\second})},
ymin=0,
ymax=150,
ytick={5,15,30,...,150},
ylabel={Energy per bit (\si[per-mode=symbol]{\pico\joule\per\bit})},
axis background/.style={fill=white},
%title={$\text Mr. Wolf)}$},
%xmajorgrids,
%xminorgrids,
%ymajorgrids,
%yminorgrids,
legend style={legend cell align=left, align=left},
legend entries={{\tiny Single SPI}, {\tiny Quad SPI SDR}, {\tiny Quad SPI DDR}, {\tiny Octal SPI SDR}, {\tiny Octal SPI DDR}, {\tiny Hyper Bus}, {\tiny SerDes} }
]

%\addplot [draw=none, mark=*, mark options={solid, gray,scale=0.25},]
%table[row sep=crcr]{%
%	1	60.0092464564725\\
%	2	58.9733678272748\\
%	4	58.834915322142\\
%	6	58.7197382901612\\
%	8	58.0382615219894\\
%	10	57.8784527510667\\
%	20	57.5928968850709\\
%	30	56.2172675778606\\
%	40	56.1847744602664\\
%	50	54.6008699582679\\
%};

%\addplot [name path=plot_spi_flash_dyn, color=color3, dashed]
%  table[row sep=crcr]{%
%0.001	58.7030009442849\\
%50		55.4478584932744\\
%};

%\addplot [draw=none, mark=*, mark options={solid, %color1,scale=0.5},]
%table[row sep=crcr]{%
%	1	62.7399033235674\\
%	2	60.3368647824497\\
%	4	59.5199914153641\\
%	6	59.1855528880754\\
%	8	58.3802578637297\\
%	10	58.1515880855453\\
%	20	57.7291743532723\\
%	30	56.3127882034048\\
%	40	56.2545254112444\\
%	50	54.6554273773945\\
%};

\addplot [name path=plot_spi_flash, mark=*, mark options={solid, black, scale=0.5}, color=black, thick]
table[row sep=crcr]{%
	0.001000	100\\
	0.050999	67.9918465334806\\
	0.100998	66.6300880126883\\
	0.150997	65.8413979202736\\
	0.250995	64.8581375818225\\
	0.350993	64.2173591104389\\
	0.450991	63.7424786216738\\
	0.600988	63.2028684470398\\
	0.750985	62.7872812858126\\
	0.900982	62.4496587459783\\
	1.100978	62.0800776162041\\
	1.300974	61.7740263318322\\
	1.550969	61.4533743678496\\
	1.800963	61.1820558181741\\
	2.100958	60.9035868652491\\
	2.400952	60.6633742227293\\
	2.750945	60.4194445639113\\
	3.150937	60.1770726612017\\
	3.600928	59.9396903091337\\
	4.100918	59.709383096814\\
	4.650907	59.4873013948615\\
	5.300894	59.2573393350504\\
	6.00088		59.0401217601795\\
	6.800864	58.8217621292076\\
	7.700846	58.6057076116804\\
	8.750825	58.384331523279\\
	9.900802	58.1712835504449\\
	11.250775	57.9515425059938\\
	12.750745	57.737193341301\\
	45.100098	55.6174446052216\\
	50	55.4478584932744\\
};

\addplot [name path=quad_spi_sdr,  mark=square*, mark options={solid, green, scale=0.5}, color=green, solid]
table[row sep=crcr]{%
0.1	75.193929 \\
0.3	39.83659566 \\
0.7	29.73450043 \\
3	23.92579566 \\
7	22.91558614 \\
10	22.688289 \\
20	22.423109 \\
30	22.33471566 \\
40	22.290519 \\
50	22.264001 \\
100	22.210965 \\
150	22.19328633 \\
200	22.184447 \\
};

\addplot [name path=quad_spi_ddr,  mark=otimes*, mark options={solid, blue, scale=0.5}, color=blue, dashed]
table[row sep=crcr]{%
0.1	69.65444675 \\
0.3	34.29711341 \\
0.7	24.19501818 \\
3	18.38631341 \\
7	17.37610389 \\
10	17.14880675 \\
20	16.88362675 \\
30	16.79523341 \\
40	16.75103675 \\
50	16.72451875 \\
100	16.67148275 \\
150	16.65380408 \\
200	16.64496475 \\
250	16.63966115 \\
300	16.63612541 \\
350	16.63359989 \\
400	16.63170575 \\
};

\addplot [name path=octalspi_sdr,  mark=diamond*, mark options={solid, cyan, scale=0.5}, color=cyan, dashed]
table[row sep=crcr]{%
0.1	96.17244675 \\
0.3	43.13644675 \\
0.7	27.98330389 \\
3	19.27024675 \\
7	17.75493246 \\
10	17.41398675 \\
20	17.01621675 \\
30	16.88362675 \\
40	16.81733175 \\
50	16.77755475 \\
100	16.69800075 \\
150	16.67148275 \\
200	16.65822375 \\
250	16.65026835 \\
300	16.64496475 \\
350	16.64117646 \\
400	16.63833525 \\
};

\addplot [name path=octalspi_ddr,  mark=star, mark options={solid, magenta, scale=0.5}, color=magenta, dashed]
table[row sep=crcr]{%
0.1	93.40270562 \\
0.3	40.36670562 \\
0.7	25.21356277 \\
3	16.50050562 \\
7	14.98519134 \\
10	14.64424562 \\
20	14.24647562 \\
30	14.11388562 \\
40	14.04759062 \\
50	14.00781362 \\
100	13.92825962 \\
150	13.90174162 \\
200	13.88848262 \\
250	13.88052722 \\
300	13.87522362 \\
350	13.87143534 \\
400	13.86859412 \\
450	13.86638429 \\
500	13.86461642 \\
550	13.86316999 \\
600	13.86196462 \\
650	13.8609447 \\
700	13.86007048 \\
750	13.85931282 \\
800	13.85864987 \\
};

\addplot [name path=ram_write512, color=color2, thick]
table[row sep=crcr]{%
0.1	113.85 \\
1600	113.85 \\
};

%\addplot [name path=ram_write512, color=color2, thick]
%table[row sep=crcr]{%
%0.1	218.4 \\
%1600	218.4 \\
%};

\addplot [name path=blink_dyn,  mark=triangle*, mark options={solid, red, scale=0.5}, color=red, thick]
table[row sep=crcr]{%
0.1	26.55114746 \\
0.3	13.21781413 \\
0.7	9.408290318 \\
1	8.551147461 \\
3	7.217814128 \\
5	6.951147461 \\
7	6.836861747 \\
10	6.751147461 \\
20	6.651147461 \\
30	6.617814128 \\
40	6.601147461 \\
50	6.591147461 \\
50	6.591147461 \\
100	6.571147461 \\
150	6.564480794 \\
200	6.561147461 \\
250	6.559147461 \\
300	6.557814128 \\
350	6.556861747 \\
400	6.556147461 \\
450	6.555591905 \\
500	6.555147461 \\
550	6.554783825 \\
600	6.554480794 \\
650	6.554224384 \\
700	6.554004604 \\
750	6.553814128 \\
793	6.553669529 \\
};

%\addplot [name path=blink_dyn_05,  mark=triangle*, mark %options={solid, red, scale=0.5}, color=red, dashed]
%table[row sep=crcr]{%
%0.1	331.1088285 \\
%0.3	115.6140368 \\
%0.6	61.74033892 \\
%1	40.19085975 \\
%3	18.64138058 \\
%5	14.33148475 \\
%7	12.48438654 \\
%10	11.09906288 \\
%20	9.482851938 \\
%30	8.944114958 \\
%40	8.674746469 \\
%50	8.513125375 \\
%100	8.189883188 \\
%150	8.082135792 \\
%200	8.028262094 \\
%250	7.995937875 \\
%300	7.974388396 \\
%350	7.958995911 \\
%400	7.947451547 \\
%450	7.938472597 \\
%500	7.931289438 \\
%516	7.929284835 \\
%};

%\node[above,color5] at (38,4) {\footnotesize{Implemented serial link}};
%\node[above,color2] at (400,31) {\footnotesize{SPI - open circuit}};
%\node[above,color1] at (40,55) {\footnotesize{SPI - S25FS256S}};

%\draw[color3] (35,65)  node[right,color3] {\footnotesize{static energy}}  -- (10,60);
%\draw[color3] (35,65)  node[right,color3] {\footnotesize{static energy}}  -- (10,60);
%\node[above,color5] at (5,1) {\scriptsize{Implemented serial link}};
%\node[above,color2] at (1,10) {\scriptsize{Quad SPI (DDR) and Octal SPI (SDR)}};
%\node[above,color4] at (-0.5,0) {\scriptsize{Octal SPI (DDR)}};
%\node[above,color6] at (-0.5,17) {\scriptsize{Quad SPI (SDR)}};
%\node[above,color1] at (2.5,54) {\scriptsize{Single SPI}};

\end{axis}
\end{tikzpicture}%
	\caption{Energy consumption compared to other peripherals (HyperBus bandwidth is fixed to 1.6Gbps)}
	\label{fig:EnergyPerBit} 
\end{figure}

Fig. \ref{fig:EnergyPerBit} shows of a comparative view of energy efficiency. We measured the energy consumption of a single SPI (in 40-nm technology) and HyperBus (in 65-nm technology), which were previously designed by our group, with an I/O voltage of 1.8V. The averaged energy of reading/writing operations were obtained. The transferred data size of HyperBus was 0.5 KB. The HyperBus is implemented by fast but power-hungry drivers, while the SPI adopts slow but low-power ones. Hence, the SPI and HyperBus operate up to 50 and 100MHz, respectively. In other words, the maximum bandwidth of the former and latter are 50 Mbps and 1.6Gbps. As shown in the graph, the HyperBus consumes much higher energy than the single SPI even though it achieves over 1Gbps. Thus, there is a trade-off between the maximum bandwidth and energy efficiency at the conventional digital interfaces.

On the other hand, the SerDes achieves a high bandwidth and low energy consumption simultaneously. Indeed, the maximum bandwidth ($BW_{max}$) with the 16KB buffer is 793Mbps. Compared to the Single SPI's best case (i.e., at 50Mbps), the SerDes efficiency is 8.46x higher at 15.9x higher performance. Besides, even if a target bandwidth is lowered to 10Mbps, the SerDes achieves 8.61x smaller energy than the SPI. Moreover, although the HyperBus achieves about two times higher bandwidth, its energy efficiency is 17.4x lower than the SerDes operating at $BW_{max}$.

Based on the SPI measurement results and its switching activity, we estimated the energy efficiency of a Quad SPI and Octal SPI operating at both DDR and SDR, shown in Fig. \ref{fig:EnergyPerBit}. As can be seen from the graph, the parallel SPI lanes improve energy efficiency. Nevertheless, the SerDes still achieves lower energy consumption. Indeed, the SerDes energy efficiency at $BW_{max}$ is 2.1x higher than the DDR Octal SPI case.

\subsection{Area overhead}
\begin{table}[t!]
	\centering
	\caption{Area overhead [$\mu{m^2}$]}
	\label{area}
	\begin{tabular}{|c|c|}
		\hline 
		SerDes & Quad SPI  \\ \hline 
		 Digital: 13427   &  \\ 
		 Analog: 30848  &  Digital: 22435 \\
		 Total: 44275 &         \\ \hline
	\end{tabular}
\end{table}

\begin{table}[t!]
	\centering
	\caption{The number of data pads needed for each solution}
	\label{number_of_pads}
	\begin{tabular}{|c|c|c|c|c|}
		\hline 
		Single SPI & Quad SPI & Octal SPI & Hyper Bus & SerDes \\ \hline 
		 4   & 6 & 11 & 12 & 6 \\ \hline 

	\end{tabular}
\end{table}

\begin{table*}[t!]
	\centering
	\caption{Comparison of this work to other short reach links recently published}
	\label{comp_table}
	\begin{tabular}{|c|c|c|c|c|c|c|}
		\hline 
		                  & Erett 2018 \cite{Erett2018} & Choi 2018 \cite{choi2018} & Poulton 2019 \cite{Poulton2019} & Shekhar 2019 \cite{Shekhar2019}& Yousry 2021 \cite{Yousry2021} & This work  \\ \hline
		  Technology      & 16-nm & 65-nm & 16-nm & 22-nm & 7-nm & 65-nm \\ \hline
	      Data Rate   & 56Gb/s/pin & 1-6Gb/s & 25Gb/s/pin & 1.2-5Gb/s & 112Gb/s & 0.8Gb/s \\ \hline 
		  Power/pin       & 126mW & 0.377-3.48mW & 28mW & 1-6.55mW& 191mW & 5.2mW \\ \hline
	  Energy efficiency   & 2.25pJ/bit & 0.29-0.58 pJ/bit & 1.17pJ/bit & 1-1.31pJ/bit & 1.7pJ/bit & 6.5pJ/bit \\ \hline
	  Supply voltage      & 0.9, 1.2, 1.8V & 0.45-0.75V & 0.85, 0.95V & 0.55-0.7V & 0.8V & 1.2V \\ \hline
	  Clocking mechanism  & CDR & Clock forwarding & Clock forwarding & CDR &CDR & CDR \\ \hline
	  Extra voltage sources & required & required & required & not required& not required & not required \\ \hline
	  System integration  & No & No & No & No &No & Yes \\ \hline

	\end{tabular}
\end{table*}

Table \ref{area} shows the area comparison excluding pads between the SerDes and the open-source Quad SPI module \cite{pulpissimo}. We obtained the value of the digital parts by area reports after the synthesis because ones after place \& route highly depend on several design choices (such as the filler cells ratio).  The area for the analog macros is the result after the custom layout. Also, Table \ref{number_of_pads} compares the number of required pads of the SerDes with other widely used digital peripherals (SPIs and HyperBus). Regarding the two chip synchronization protocol in Section \ref{sec:chips_sync}, additional two GPIOs are necessary for the SerDes, resulting in six pins in total.

As shown in Table \ref{area}, the SerDes consumes more area than the Quad SPI module as it requires analog macros occupied by large passive components and wider transistors. So in case that a required silicon footprint is limited, the full digital interface can be an option. However, regarding Table \ref{number_of_pads} and Fig. \ref{fig:EnergyPerBit}, the high bandwidth with pure digital interfaces is achieved at the cost of pad counts and energy consumption, critical for microcontrollers driven by batteries and equipping a few pins. On the other hand, the SerDes interface needs less or the same number of pads than the others and offers the smallest energy in Fig. \ref{fig:EnergyPerBit}.

\subsection{Comparison to other short reach links}
Table \ref{comp_table} summarizes a comparison to other short reach links recently published. While the other links achieve higher bandwidth and energy-efficiency than ours, they are missing the option to be integrated into actual microcontrollers. Indeed, the works from \cite{Erett2018,Poulton2019,Yousry2021} are not suitable for low power applications as their total power consumption is much higher than the power envelope of an IoT edge devices (10mW). Also, the work in \cite{choi2018} does not consider power overhead imposed by extra voltage sources. Moreover, the clock forwarding scheme requires additional pad resources. The work from Shekhar et al. \cite{Shekhar2019} addresses these problems; however, this design is still a stand-alone system like the ones in  \cite{Erett2018,Poulton2019,choi2018} and ignores the essential functionality required by actual microcontrollers. On the other hand, our design meets all the requirements, such as the tight power envelope, limited number of pads, and system integration. Furthermore, our power efficiency is still much higher than other pure digital peripherals.

\section{Conclusion} \label{sec:conc}
This paper presented the first complete SoC design implementing a high-speed serial link for IoT edge devices and evaluated the energy efficiency based on real chips fabricated with 65-nm CMOS technology. The SerDes achieves lower energy consumption and higher bandwidth of the data communication than existing digital peripheral interfaces such as parallel SPIs and HyperBus, let alone a single SPI. Indeed, it records about two times less energy than the octal SPI interface at the same data rate target. Moreover, we also reveal that the SerDes saves the number of required pads which is often beneficial for embedded systems.

In the current work, the analog parts' power consumption is not fully optimized because this is the first prototype. Indeed, 80\% of the power is dissipated by them, resulting in relatively higher energy than other short-range serial links. Also, our SerDes does still not employ the near-threshold voltage region, unlike other low-power links. Consequently, there is still room to lower the power consumption further. Addressing these concerns is our main future work.

\begin{IEEEbiography}[{\includegraphics[width=1in,clip,keepaspectratio]{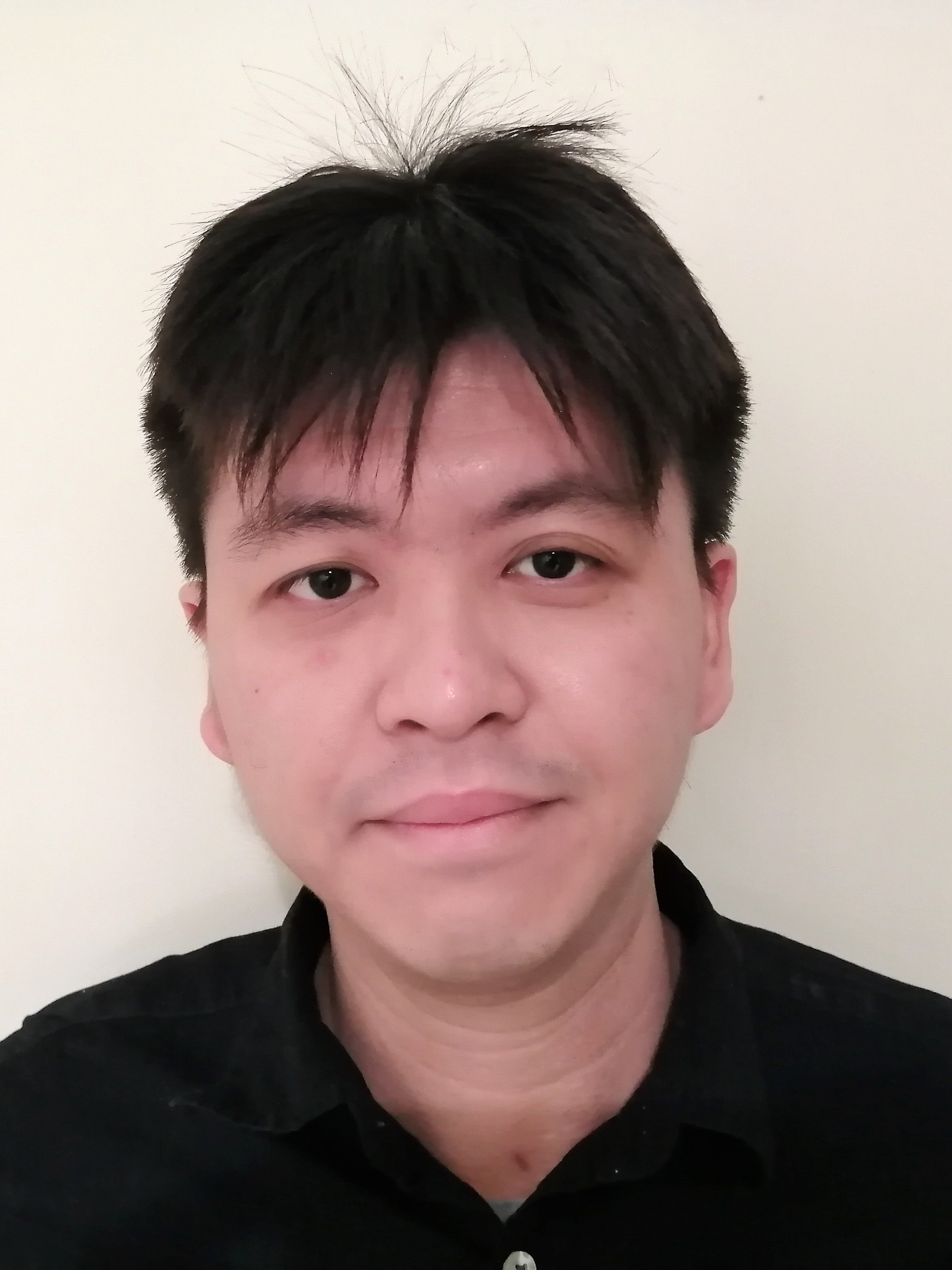}}]{Hayate Okuhara}
%\begin{IEEEbiography}{Hayate Okuhara}
 received the Ph.D. degree from Keio University, Japan, in 2018. He was a Postdoctoral Researcher with the Department of Electrical, Electronic and Information Engineering “Guglielmo Marconi,” University of Bologna, Italy. His research interest includes low-power VLSI systems such as energy-efficient serial interfaces.
\end{IEEEbiography}

\begin{IEEEbiography}[{\includegraphics[width=1in,clip]{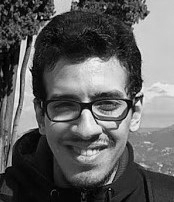}}]{Ahmed Elnaqib}
%\begin{IEEEbiography}{Ahmed Elnaqib}
received his Bachelor's degree in electronic engineering from Ain Shams University, Cairo, Egypt in 2014. He received his Master's degree from the University of Bologna, Italy in 2018. From 2018 to 2020, he was with the University of Bologna, where he served as a researcher in low-power high-speed serial links. Currently, he is with Caeleste developing custom CMOS image sensors.
\end{IEEEbiography}

\begin{IEEEbiography}[{\includegraphics[width=1in,clip]{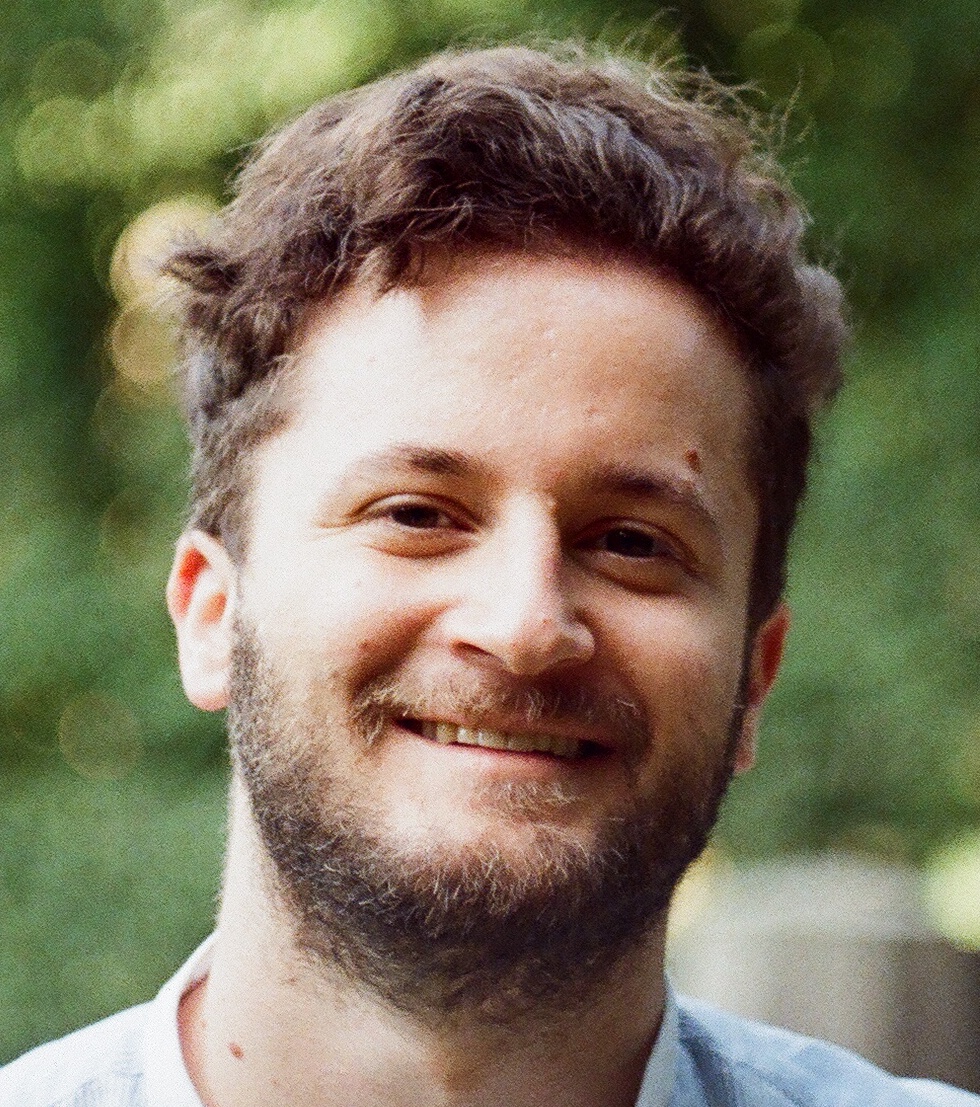}}]{Martino Dazzi}
%\begin{IEEEbiography}{Hayate Okuhara}
received his BSc (2015) and MSc (2017, cum laude) in Electronic Engineering from University of Udine, Italy. From May 2017 to January 2018 he held a Research Assistant position at Alma Mater Studiorum - Universita’ di Bologna, Italy. Since 2018, he is pursuing his PhD as a pre-doctoral researcher at the Neuropmorphic Computing and I/O links group at IBM
Research-Zurich, in conjunction with ETH Zurich. His main research interest is in energy-efficient design of neural network accelerators, specifically in reduced precision implementations and the computational memory based-, non-von Neumann computing paradigm.
\end{IEEEbiography}

\begin{IEEEbiography}[{\includegraphics[width=1in,clip]{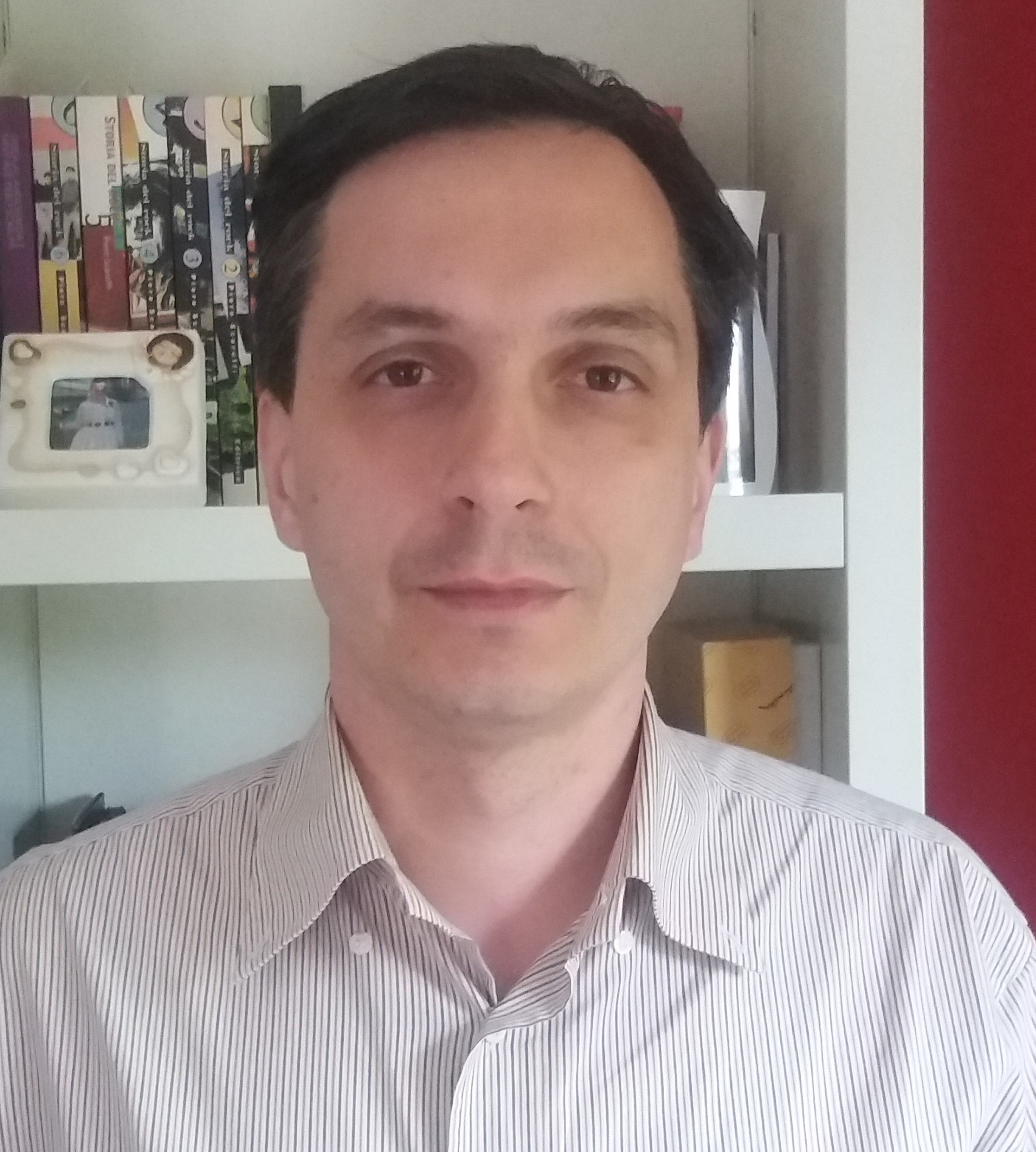}}]{Pierpaolo Palestri}
received the Laurea Degree in Electronic Engineering from the University of Bologna, Italy, in 1998, and the Ph.D. in Electronic Engineering from the University of Udine, Italy, in 2003. In 2000 and 2001 he held a post-doctoral position at Bell Laboratories, Murray Hill, NJ. In 2005 he became Associate Professor at the University of Udine.

His research interests include the modeling of carrier transport in nanoscale devices, as well as the design of integrated circuits for high-frequency applications. He has coauthored about 300 papers in international journals and conferences with peer review.
\end{IEEEbiography}

\begin{IEEEbiography}[{\includegraphics[width=1in,height=1.25in,clip,keepaspectratio]{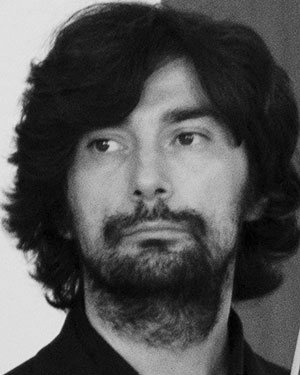}}]{Simone Benatti}
received the Ph.D. degree in electrical engineering and computer science from the University of Bologna, Bologna, Italy, in 2016. He has collaborated with several international research institutes and companies. Previously, he worked for 8 years as an Electronic Designer and R\&D Engineer of electromedical devices. In this ﬁeld, he has authored or coauthored more than 50 papers in international peer-reviewed conferences and journals. His research interests focus on energy efﬁcient embedded wearable systems, signal processing, sensor fusion, and actuation systems. This includes hardware/software codesign to efﬁciently address performance, as well as advanced algorithms.
\end{IEEEbiography}

\begin{IEEEbiography}[{\includegraphics[width=1in,height=1.25in,clip,keepaspectratio]{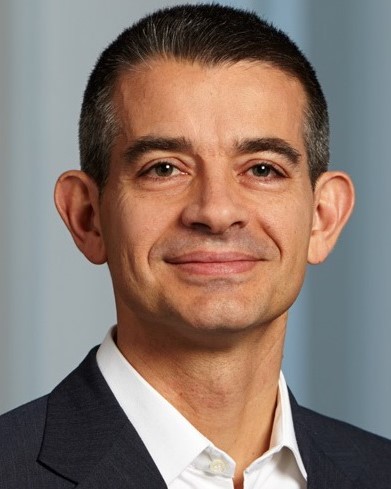}}]{Luca Benini}
holds the chair of digital Circuits and systems at ETHZ and is Full Professor at the University of Bologna. Dr. Benini’s research interests are in energy-efficient computing systems design, from embedded to highperformance. He has published more than 1000 peer-reviewed papers and five books. He is a Fellow of the ACM and a member of the Academia Europaea. He is the recipient of the 2016 IEEE CAS Mac Van Valkenburg Award and the 2020 EDAA Achievement Award.
\end{IEEEbiography}

\begin{IEEEbiography}[{\includegraphics[width=1in,clip]{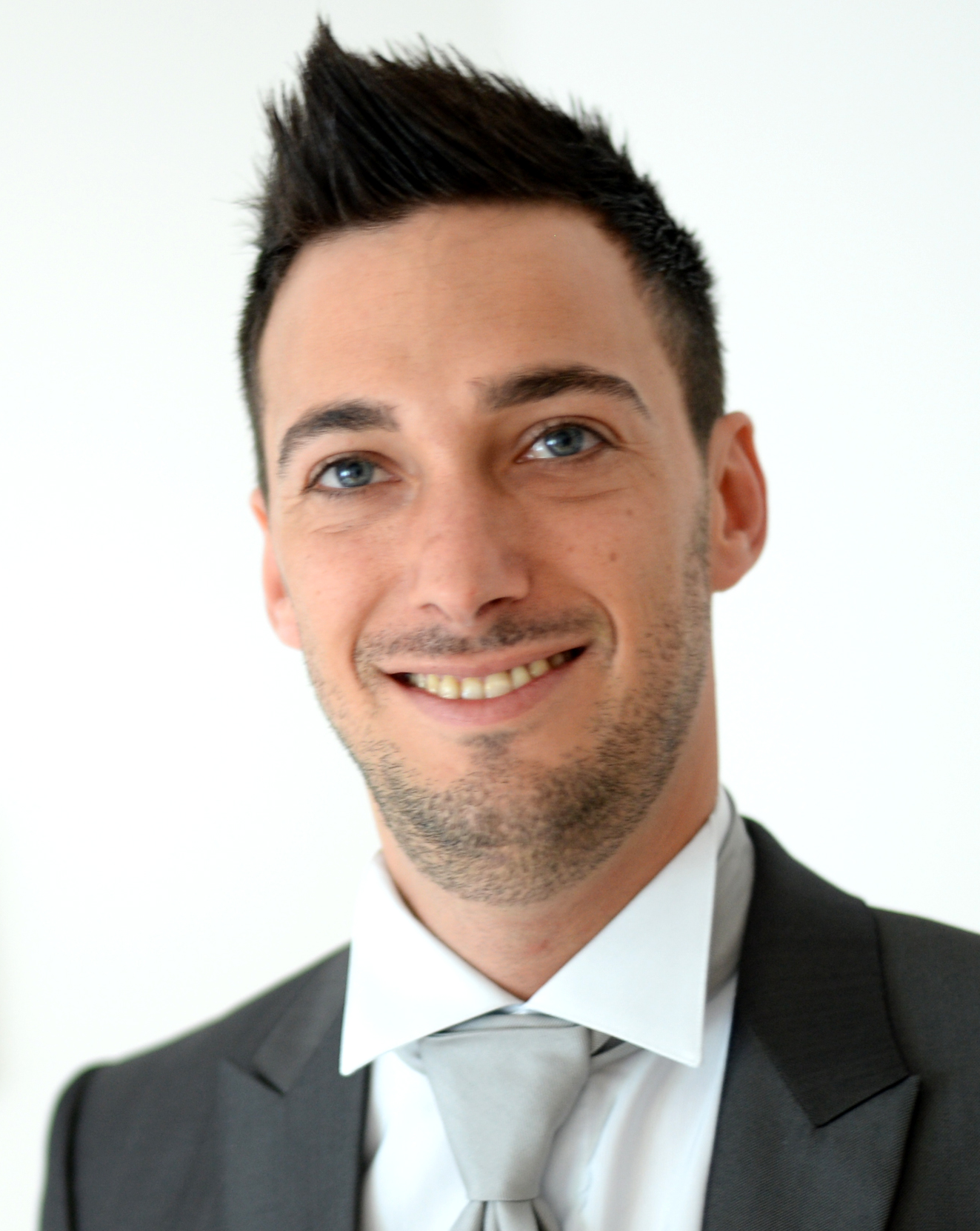}}]{Davide Rossi}
received the PhD from the University of Bologna, Italy, in 2012 where he currently holds an associate professor position. His research interests focus on energy efficient digital architectures in the domain of heterogeneous and reconfigurable multi and many-core systems on a chip. This includes architectures, design implementation strategies, and runtime support to address performance, energy efficiency, and reliability issues of both high end embedded platforms and ultra-low-power computing platforms targeting the IoT domain. In these fields he has published more than 100 papers in international peer-reviewed conferences and journals. He is recipient of Donald O. Pederson Best Paper Award 2018, - 2020 IEEE Transactions on Circuits and Systems Darlington Best Paper Award, 2020 IEEE Transactions on Very Large Scale Integration Systems Prize Paper Award.
\end{IEEEbiography}


\begin{thebibliography}{00}
\bibitem{samurAI}  I. Miro-Panades et al., "SamurAI: A 1.7MOPS-36GOPS Adaptive Versatile IoT Node with 15,000× Peak-to-Idle Power Reduction, 207ns Wake-Up Time and 1.3TOPS/W ML Efficiency," 2020 IEEE Symposium on VLSI Circuits, Honolulu, HI, USA, 2020, pp. 1-2.
\bibitem{mrwolf} A. Pullini, D. Rossi, I. Loi, G. Tagliavini and L. Benini, "Mr.Wolf: An Energy-Precision Scalable Parallel Ultra Low Power SoC for IoT Edge Processing," in IEEE Journal of Solid-State Circuits, vol. 54, no. 7, pp. 1970-1981, July 2019.
\bibitem{vega} D. Rossi et al., "4.4 A 1.3TOPS/W @ 32GOPS Fully Integrated 10-Core SoC for IoT End-Nodes with 1.7$\mu$W Cognitive Wake-Up From MRAM-Based State-Retentive Sleep Mode," 2021 IEEE International Solid- State Circuits Conference (ISSCC), San Francisco, CA, USA, 2021, pp. 60-62.
\bibitem{esweek2019} A. Burrello, F. Conti, A. Garofalo, D. Rossi, and L. Benini, "Work-in-Progress: DORY: Lightweight Memory HierarchyManagement for Deep NN Inference on IoT Endnodes", International Conference on Hardware/Software Codesign and System Synthesis (CODES+ISSS 2019), pp.1-2.
\bibitem{MobileNet} Sandler, Mark et al., "Mobilenetv2: Inverted residuals and linear bottlenecks", 2018 IEEE/CVF Conference on Computer Vision and Pattern Recognition, Salt Lake City, UT, 2018, pp. 4510--4520.
\bibitem{apmemory} AP Memory Technology Corp., IoT RAM \& ADMUX PSRAM. Available Online:
http://www.apmemory.com/html/product$\_$psram.php. (last checked:09/05/2019 )
\bibitem{hyperRAM} Cypress Semiconductors, HyperRAM Memory, Available Online:
http://www.cypress.com/products/hyperram-memory, (last checked 07/30/2019).
\bibitem{ssi} Synopsys, DesignWare Synchronous Serial Interface (SSI), Available Online:
https://www.synopsys.com/dw/ipdir.php?ds=amba$\_$ssi. (last checked: 12/03/2021)
\bibitem{synopsys_ip} Synopsys, DESIGNWARE IP DATASHEET USR/XSR PHY, Available Online:
https://www.synopsys.com/dw/doc.php/ds/c/dwc$\_$USR-XSR-die-to-die$\_$phy$\_$ds.pdf. (last checked:13/03/2021)
\bibitem{nearth} R. G. Dreslinski, M. Wieckowski, D. Blaauw, D. Sylvester and T. Mudge, "Near-Threshold Computing: Reclaiming Moore's Law Through Energy Efficient Integrated Circuits," in Proceedings of the IEEE, vol. 98, no. 2, pp. 253-266, Feb. 2010.
\bibitem{manticore}F. Zaruba, F. Schuiki and L. Benini, "Manticore: A 4096-core RISC-V Chiplet Architecture for Ultra-efficient Floating-point Computing," in IEEE Micro (Early Access).
\bibitem{lvds} Texas instruments, "Interface circuits for TIA/EIA-644 (LVDS)", Available online: https://www.ti.com/lit/an/slla038b/slla038b.pdf (last checked 17/11/2020).
\bibitem{pulpissimo} PULP-platform, PULPissimo, Available Online:
https://github.com/pulp-platform/pulpissimo, (last checked 07/30/2019).
\bibitem{Yang1996} Chih-Kong Ken Yang and M. A. Horowitz, "A 0.8-/spl mu/m CMOS 2.5 Gb/s oversampling receiver and transmitter for serial links," in IEEE Journal of Solid-State Circuits, vol. 31, no. 12, pp. 2015-2023, Dec. 1996.
\bibitem{Dally1997} William J. Dally and John Poulton, "Transmitter Equalization for 4-GBPS Signaling," IEEE Micro 17, 1, pp. 48–56, Jan. 1997.
\bibitem{Lee2000} M. -. E. Lee, W. J. Dally and P. Chiang, "Low-power area-efficient high-speed I/O circuit techniques," in IEEE Journal of Solid-State Circuits, vol. 35, no. 11, pp. 1591-1599, Nov. 2000.
\bibitem{Kim2002}J. Kim and M. A. Horowitz, "Adaptive supply serial links with sub-1-V operation and per-pin clock recovery," in IEEE Journal of Solid-State Circuits, vol. 37, no. 11, pp. 1403-1413, Nov. 2002.
\bibitem{Bulzacchelli2006} J. F. Bulzacchelli et al., "A 10-Gb/s 5-Tap DFE/4-Tap FFE Transceiver in 90-nm CMOS Technology," in IEEE Journal of Solid-State Circuits, vol. 41, no. 12, pp. 2885-2900, Dec. 2006.
\bibitem{Poulton2013}J. W. Poulton et al., "A 0.54 pJ/b 20 Gb/s Ground-Referenced Single-Ended Short-Reach Serial Link in 28 nm CMOS for Advanced Packaging Applications," in IEEE Journal of Solid-State Circuits, vol. 48, no. 12, pp. 3206-3218, Dec. 2013.
\bibitem{choi2015} W. Choi et al., "3.8 A 0.45-to-0.7V 1-to-6Gb/S 0.29-to-0.58pJ/b source-synchronous transceiver using automatic phase calibration in 65nm CMOS," 2015 IEEE International Solid-State Circuits Conference - (ISSCC) Digest of Technical Papers, San Francisco, CA, 2015, pp. 1-3.
\bibitem{choi2018} W. Choi et al., "A 0.45-0.7 V 1-6 Gb/s 0.29-0.58 pJ/b Source-Synchronous Transceiver Using Near-Threshold Operation," in IEEE Journal of Solid-State Circuits, vol. 53, no. 3, pp. 884-895, March 2018. 
\bibitem{dazzi2018} M. Dazzi et al., "Sub-mW multi-Gbps chip-to-chip communication Links for Ultra-Low Power IoT end-nodes," 2018 IEEE International Symposium on Circuits and Systems (ISCAS), Florence, 2018, pp. 1-5.
\bibitem{Poulton2019} J. W. Poulton et al., "A 1.17-pJ/b, 25-Gb/s/pin Ground-Referenced Single-Ended Serial Link for Off- and On-Package Communication Using a Process- and Temperature-Adaptive Voltage Regulator," in IEEE Journal of Solid-State Circuits, vol. 54, no. 1, pp. 43-54, Jan. 2019.
\bibitem{Pisati2020} M. Pisati et al., "A 243-mW 1.25–56-Gb/s Continuous Range PAM-4 42.5-dB IL ADC/DAC-Based Transceiver in 7-nm FinFET," in IEEE Journal of Solid-State Circuits, vol. 55, no. 1, pp. 6-18, Jan. 2020.
\bibitem{Yousry2021} R. Yousry et al., "11.1 A 1.7pJ/b 112Gb/s XSR Transceiver for Intra-Package Communication in 7nm FinFET Technology," 2021 IEEE International Solid- State Circuits Conference (ISSCC), San Francisco, CA, USA, 2021, pp. 180-182.
\bibitem{Shekhar2019}S. Shekhar, R. Inti, J. Jaussi, T. Hsueh and B. K. Casper, "A Low-Power Bidirectional Link With a Direct Data-Sequencing Blind Oversampling CDR," in IEEE Journal of Solid-State Circuits, vol. 54, no. 6, pp. 1669-1681, June 2019
\bibitem{Erett2018}M. Erett et al., "A 126mW 56Gb/s NRZ wireline transceiver for synchronous short-reach applications in 16nm FinFET," 2018 IEEE International Solid - State Circuits Conference - (ISSCC), San Francisco, CA, pp. 274-276, 2018.
\bibitem{wright2020}J. C. Wright et al., "A Dual-Core RISC-V Vector Processor With On-Chip Fine-Grain Power Management in 28-nm FD-SOI," in IEEE Transactions on Very Large Scale Integration (VLSI) Systems (Early Access).
\bibitem{Amid2020}C. Schmidt et al., "Programmable Fine-Grained Power Management and System Analysis of RISC-V Vector Processors in 28-nm FD-SOI," in IEEE Solid-State Circuits Letters, vol. 3, pp. 210-213, 2020.
\bibitem{Okuhara2020} H. Okuhara et al., "An Energy-Efficient Low-Voltage Swing Transceiver for mW-Range IoT End-Nodes," 2020 IEEE International Symposium on Circuits and Systems (ISCAS), Sevilla, 2020, pp. 1-5.
\bibitem{alexander1975} J. Alexander, “Clock recovery from random binary signals,” Electron.
Lett., vol. 11, no. 22, pp. 541–542, 1975.
\bibitem{amd_chiplet}S. Naffziger, K. Lepak, M. Paraschou and M. Subramony, "2.2 AMD Chiplet Architecture for High-Performance Server and Desktop Products," 2020 IEEE International Solid- State Circuits Conference - (ISSCC), San Francisco, CA, USA, 2020, pp. 44-45.
\bibitem{razavi2002} B. Razavi, "Challenges in the design high-speed clock and data recovery circuits," in IEEE Communications Magazine, vol. 40, no. 8, pp. 94-101, Aug. 2002.
\bibitem{blackghost} Y. Pu et al., "A 9-mm2 Ultra-Low-Power Highly Integrated 28-nm CMOS SoC for Internet of Things," in IEEE Journal of Solid-State Circuits, vol. 53, no. 3, pp. 936-948, March 2018.
\bibitem{intel2018}  T. Karnik et al., "A cm-scale self-powered intelligent and secure IoT edge mote featuring an ultra-low-power SoC in 14nm tri-gate CMOS," 2018 IEEE International Solid - State Circuits Conference - (ISSCC), San Francisco, CA, 2018, pp. 46-48.
\bibitem{udma} A. Pullini, D. Rossi, G. Haugou and L. Benini, "$\mu$DMA: An autonomous I/O subsystem for IoT end-nodes," 2017 27th International Symposium on Power and Timing Modeling, Optimization and Simulation (PATMOS), Thessaloniki, 2017, pp. 1-8.
\bibitem{8b10b}A. X. Widmer, P. A. Franaszek, “A DC-balanced, partitioned-block, 8B/10B transmission code”, IBM Journal of Research and Development, v.27 n.5, p.440-451, September 1983.
\bibitem{comparator}S. Babayan-Mashhadi and R. Lotfi, "Analysis and Design of a Low-Voltage Low-Power Double-Tail Comparator," in IEEE Transactions on VLSI Systems, vol. 22, no. 2, pp. 343-352, Feb. 2014.
\bibitem{elnaqib2020}A. Elnaqib, H. Okuhara, T. Jang, D. Rossi and L. Benini, "A 0.5GHz 0.35mW LDO-Powered Constant-Slope Phase Interpolator with 0.22\% INL," in IEEE Transactions on Circuits and Systems II: Express Briefs (Early access).
\bibitem{Usami2011} K. Usami et al., "On-chip detection methodology for break-even time of power gated function units," IEEE/ACM International Symposium on Low Power Electronics and Design, Fukuoka, Japan, 2011, pp. 241-246.
\bibitem{Anand2014}T. Anand, A. Elshazly, M. Talegaonkar, B. Young and P. K. Hanumolu, "A 5 Gb/s, 10 ns Power-On-Time, 36 $\mu$W Off-State Power, Fast Power-On Transmitter for Energy Proportional Links," in IEEE Journal of Solid-State Circuits, vol. 49, no. 10, pp. 2243-2258, Oct. 2014.
\bibitem{umc65} UMC, 65nm technology. Available Online:
http://www.umc.com/english/pdf/UMC\%2065nm.pdf, June 2005. (last checked:09/05/2019)
\bibitem{pcb_calc} Coupled Microstrip, Analysis/Synthesis Calculator. Available Online: http://wcalc.sourceforge.net/cgi-bin/coupled$\_$microstrip.cgi, 2009. (last checked:06/02/2021 )
\bibitem{FLL} D. Bellasi, P. Schönle, Q. Huang and L. Benini, "A wide tuning-range ADFLL for mW-SoCs with dithering-enhanced accuracy in 65 nm CMOS," 2017 IEEE International Symposium on Circuits and Systems (ISCAS), Baltimore, MD, 2017, pp. 1-4.
\bibitem{lvds-product} Analog Devices, 3V LVDS Quad CMOS Differential Line Driver ADN4665. Available Online:https://www.analog.com/media/en/technical-documentation/data-sheets/ADN4665.pdf (last checked: 30/06/2021)
\end{thebibliography}
\end{document}